\documentclass[reprint,prl,aps,amsmath,amssymb,floatfix,superscriptaddress,longbibliography,cite]{revtex4-2}

 \usepackage{mathtools,amsmath}
\usepackage{graphicx} 
\usepackage{tikz}
\usepackage[breaklinks=true,colorlinks,citecolor=teal,linkcolor=teal,urlcolor=teal]{hyperref}
\usepackage{physics}
\usepackage{siunitx}
\usepackage{bbold}
\usepackage{soul}
\usepackage{bm}
\usepackage[normalem]{ulem}
\usepackage{times}
\usepackage{enumerate}  
\usepackage{float}
\usepackage{amssymb}
\usepackage{xcolor} 
\usepackage{bibunits}

\renewcommand{\Re}{\mathop{\text{Re}}\nolimits}

\def\la{\langle}
\def\ra{\rangle}

\def\be{\begin{equation}}
\def\ee{\end{equation}}

\begin{document}

\title{Universal Characterization of Quantum Vacuum Measurement Engines}

\author{Robert Czupryniak}
\affiliation{Department of Physics and Astronomy, University of Rochester, Rochester, NY 14627, USA}
\affiliation{Institute for Quantum Studies, Chapman University, Orange, CA 92866, USA}
\author{Bibek Bhandari}
\affiliation{Institute for Quantum Studies, Chapman University, Orange, CA 92866, USA}
\author{Paolo Andrea Erdman}
\affiliation{Department of Mathematics and Computer Science, Freie Universit\"at Berlin, 14195 Berlin, Germany}
\author{Andrew N Jordan}
\affiliation{The Kennedy Chair in Physics, Chapman University, Orange, CA 92866, USA}
\affiliation{Schmid College of Science and Technology, Chapman University, Orange, CA 92866, USA}
\affiliation{Institute for Quantum Studies, Chapman University, Orange, CA 92866, USA}
\affiliation{Department of Physics and Astronomy, University of Rochester, Rochester, NY 14627, USA}

\begin{abstract}
Quantum measurements can inject energy into quantum systems, enabling engines whose operation is powered entirely by measurements. We develop a general theory of quantum vacuum measurement engines by introducing the {\em quantum vacuum bending function} (QVBF), a quantity that characterizes the lowering of the ground-state energy due to interactions. We show that all thermodynamic observables, including work and efficiency, are governed solely by the shape of the ground-state energy landscape encoded in the QVBF, regardless of microscopic details.  We further demonstrate that work fluctuations are defined by the curvature of QVBF modulated by a model-dependent quantity, and are constrained by a generalized quantum fluctuation relation that involves the interplay between quantum Fisher information and the ground-state energy landscape. Exactly solvable models and numerical simulations of single and many-body systems confirm the theory and illustrate how the QVBF alone determines the performance of quantum vacuum measurement engines.

\end{abstract}

\maketitle
The interplay between quantum measurement and thermodynamics has attracted growing interest in recent years. Unlike classical observations, quantum measurement can alter the state of the system \cite{jordan2024quantum}, injecting energy \cite{elouard2017, latune2025thermodynamically,nath2025capturing} and creating a non-trivial flow of information \cite{yanik2022thermodynamics}. This backaction has been shown to enable measurement-powered thermodynamic machines, whose functionality does not rely on externally supplied work but only on the energy provided by the measurement itself \cite{elouard2017-1,elouard2017,yi2017,mohammady2017,elouard2018,ding2018,jordan2020,elouard2020interaction,bresque2021,manikandan2022,jussiau2023many,das2023quantum,sanchez2025making,elouard2025revealing, erdman2025artificially, bhandari2023measurement, buffoni2019quantum,continuous_bibek}. 
The quantum vacuum measurement engine exploits measurements applied on a many-body system prepared in its ground state and allows work extraction even at zero temperature \cite{jordan2004entanglement,
jussiau2023many}.  Research into energetic uses of the quantum vacuum has been an ongoing topic of investigation \cite{forward1984extracting,cole1993extracting,PhysRevB.92.125306,hofer2016quantum}.
\begin{figure}[t]
    \centering
    \includegraphics[width=1.\linewidth]{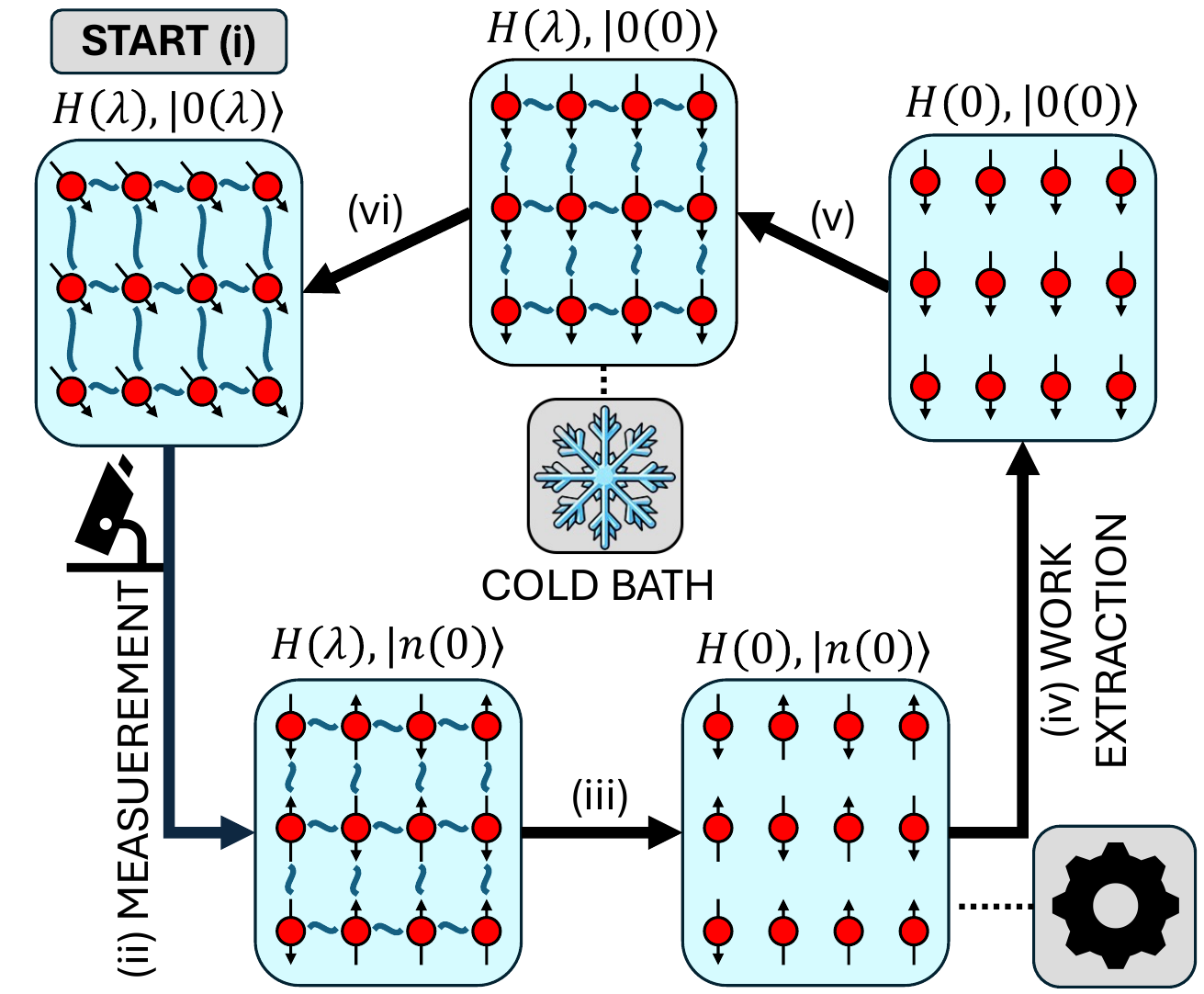}
    \caption{Schematic of a quantum vacuum measurement engine in a many-body setting. Switching on the parameter $\lambda$ corresponds to activating the coupling between spins. The labels above each square indicate whether the coupling in the Hamiltonian is on ($H(\lambda)$) or off ($H(0)$), followed by the quantum state of the system at that stage of the cycle. The protocol consists of six steps: (i) initialization, (ii) measurement, (iii) rapid switching off of the coupling, (iv) work extraction, (v) rapid switching on of the coupling, and (vi) relaxation.
    }
    \label{fig:sketch}
\end{figure}

Many-body systems can exhibit complex ground-state energetics that are sensitive to interactions and correlations, making them powerful platforms for quantum thermodynamic tasks \cite{dowling2004energy,jussiau2023many}. In certain regimes, the ground-state structure itself (shaped by many-body interactions) determines the system’s ability to absorb and release energy, particularly in the form of phase transitions \cite{sachdev1999quantum}. Jussiau {\it et al}.~\cite{jussiau2023many} introduced a class of vacuum measurement engines where the thermodynamic output is fully governed by properties of the interacting ground state. However, a broadly applicable theoretical perspective that identifies and unifies the ground-state features relevant for thermodynamic performance across different physical platforms has yet to be fully developed. Establishing such a perspective would help clarify the fundamental thermodynamic role of many-body structure and provide guidance for designing quantum engines beyond model-specific analyses.

In this Letter, we demonstrate that the thermodynamic behavior of vacuum measurement engines is entirely governed by the \textit{quantum vacuum bending function} (QVBF) and its derivative with respect to a parameter controlling the coupling strength. 
The QVBF is defined as how the ground state energy changes as a function of a Hamiltonian parameter, in analogy to ``band bending'' in semiconductor physics \cite{mott1938note}.
The QVBF generalizes the concept of \textit{local entanglement gap} introduced in Ref.~\cite{jussiau2023many}, allowing the description of systems with arbitrary coupling types beyond pairwise interactions. Jordan and B\"uttiker pointed out that measurements in the local energy basis can yield excited energies, serving also as an entanglement witness \cite{jordan2004entanglement}.
We further demonstrate that this geometric description applies universally to both single and many-body systems, revealing a common underlying structure. We find that work fluctuations are set by the curvature of the QVBF, establishing a direct connection between thermodynamic noise and the geometry of the ground-state energy landscape. Building on this framework, we derive an effective thermodynamic uncertainty relation that bounds work fluctuations in terms of the quantum Fisher information. By expressing the quantum Fisher information in terms of experimentally accessible quantities such as work and its fluctuations, our results also provide a practical route to making fundamental bounds on quantum metrological sensitivity. 

Together, these findings establish a universal and experimentally relevant characterization of vacuum measurement engines. To illustrate the generality of our approach, we apply the framework to representative examples based on qubits and harmonic oscillators, demonstrating how the QVBF geometry alone determines the work, efficiency, and fluctuations of measurement-powered engines across distinct physical platforms.

\textit{Quantum Vacuum Measurement Engines.}---We consider a general quantum system serving as the working medium, described by a Hamiltonian decomposed into two parts
\begin{equation}\label{eq:h_lambda}
    H(\lambda)=H_\text{loc}+\lambda H_\text{int}.
\end{equation}
The \textit{local part}, $H_\text{loc}$, defines the measurement basis and may correspond to either a single quantum system, such as a qubit or a harmonic oscillator, or a collection of noninteracting subsystems in a many-body setting. The \textit{interaction part} $H_\text{int}$ captures processes that enable engine's operation, e.g., transverse driving in single-system scenarios or entangling interactions in many-body systems. 
We introduce a tunable parameter—the \textit{coupling strength} $\lambda$—which controls the strength of the interaction term and specifies the operating point of the engine when the interaction is switched on.
We note that our framework can be applied to single-body, non-interacting systems, and the naming convention for $H_\text{loc}$ and $H_\text{int}$ is chosen to keep the notation consistent with Ref.~\cite{jussiau2023many}. 

Let $|n(\lambda)\rangle$ and $E_n(\lambda)$ denote the eigentstates of $H(\lambda)$. Setting $\lambda=0$ recovers the eigenbasis of the uncoupled system. We impose the condition $\langle n(0)|H_\text{int}|n(0)\rangle=0$, so that the interactions can be abruptly switched on and off at no energetic cost if the system is in the state $|n(0)\rangle$. We assume $E_0(\lambda)\leq E_0(0)$.

As shown in Fig.~\ref{fig:sketch}, the engine's cycle consists of six steps: (i) with coupling on and Hamiltonian $H(\lambda)$, we start with the quantum system in ground state $|0(\lambda)\rangle$; (ii) perform projective measurement in the eigenbasis of $H_\text{loc}$; (iii) switch off the coupling, $H(\lambda)\rightarrow H(0)$; (iv) work extraction: coherently rotate the system to the ground state of the local part $H_\text{loc}$, $|0(0)\rangle$; (v) switch coupling back on, $H(0)\rightarrow H(\lambda)$; (vi) allow the system to relax to $|0(\lambda)\rangle$ by contacting with a zero-temperature bath. The cycle is then repeated. 

In this letter, we assume a non-degenerate ground state $|0(\lambda)\rangle$. The expected work per cycle is $
    W(\lambda)=\sum_n |\la 0(\lambda)|n(0) \ra|^2 (E_n(0)-E_0(0)) = \langle H_\text{loc}\rangle_{\lambda} - E_0(0),$
where $\{ E_n(0), |n(0)\ra\} $ are the energy eigenvalues and eigenstates of the local systems, and the expectation is defined as $\langle \bullet \rangle_{\lambda} = \langle 0(\lambda)|\bullet|0(\lambda)\rangle$ \cite{jussiau2023many}. Quantum heat~\cite{auffeves2022} is defined as the average energy injected into the system by the measurement apparatus per cycle,
$Q(\lambda)
= -\lambda \langle H_\text{int} \rangle_{\lambda}$.
The engine's efficiency is $
    \eta(\lambda) = W(\lambda) / Q(\lambda),$
and the fluctuations of the engines work per cycle are $\sigma^2(\lambda) = \langle H_\text{loc}^2\rangle_{\lambda}- \langle H_\text{loc} \rangle_{\lambda}^2.$

\textit{Geometric framework.}---The above relations for work, quantum heat, efficiency, and fluctuations were derived in Ref.~\cite{jussiau2023many} for a specific choice of $H_\text{loc}$ and $H_\text{int}$. We define the \textit{quantum vacuum bending function}
\begin{equation}
\label{eq:delta_etienne}
\Delta(\lambda) = E_0(0) - E_0(\lambda),
\end{equation}
as the coupling-induced lowering of the ground-state energy. By treating $\lambda$ as a tunable parameter, we show below that all thermodynamic quantities, $W(\lambda)$, $Q(\lambda)$, and $\eta(\lambda)$, depend only on the QVBF and its derivative, while the fluctuations depend on
its curvature up to a model-dependent function. This identifies $\Delta(\lambda)$ as the central geometric quantity characterizing the performance of quantum vacuum measurement engines.

To connect the engine's performance to the QVBF, we use the Hellmann-Feynman theorem \cite{ hellmann1937einfuhrung, feynman1939forces} to write $E_0'(\lambda)=\langle H'(\lambda)\rangle_{\lambda}=\langle H_\text{int}\rangle_{\lambda}$, where the prime symbol $(')$ denotes the derivative $d/d\lambda$. Since $\Delta'(\lambda)=-E_0'(\lambda)$, the quantum heat becomes
\begin{equation} \label{eq:q_lambda}
    Q(\lambda) = \lambda\Delta'(\lambda),
\end{equation}
demonstrating that the quantum heat is governed by the slope of the QVBF. Similarly, the average work and efficiency can be expressed as
\begin{equation} \label{eq:W_eta_lambda}
    W(\lambda) = \lambda \Delta'(\lambda)-\Delta(\lambda), \quad \eta(\lambda)=1-\frac{\Delta(\lambda)}{\lambda\Delta'(\lambda)},
\end{equation}
showing that {\em the thermodynamic properties of a quantum measurement engine depend only on $\Delta(\lambda)$ and its slope}.  We stress the nonperturbative nature of this result. For any operational engine $W(\lambda)>0\implies\lambda \Delta'(\lambda)>\Delta(\lambda)$, which enforces $\eta(\lambda)$ to be strictly below $1$.

Work fluctuations are governed by the curvature of the QVBF. As shown in the Supplemental Material~\cite{SM},
\begin{equation}\label{eq:sigma_lambda}
    \sigma^2(\lambda) = \frac{1}{2}\lambda^2 \Delta''(\lambda) \bar{e}(\lambda),
\end{equation}
where $\bar{e}(\lambda)$ is a funtion determined by matrix elements of $H_\text{int}$ between the ground and excited states of $H(\lambda)$. It is defined through
${1}/{\bar{e}} \equiv {\sum_{n>0} (c_n^2/e_n)}/{\sum_{n>0} c_n^2}$,
with $c_n^{2} \equiv |\langle n(\lambda)|H_\text{int}|0(\lambda)\rangle|^{2}$ and $e_n \equiv E_n(\lambda)-E_0(\lambda)$. Since $\bar{e}(\lambda)$ is a harmonic mean weighted by the coefficients $c_n^2$, it is bounded between the minimum and maximum excitation energies accessible through $H_\text{int}$. Denoting these bounds by $\bar{e}_{\mathrm{min}}$ and $\bar{e}_{\mathrm{max}}$, respectively, we obtain the fundamental bounds
\begin{equation} \label{eq:fluctuation_bounds}
    \frac{\lambda^2 \Delta''(\lambda) \bar{e}_{\text{min}}}{2}\leq \sigma^2(\lambda) \leq \frac{\lambda^2 \Delta''(\lambda) \bar{e}_{\text{max}}}{2}.
\end{equation}
An upper bound exists only when the local subsystems possess a bounded energy spectrum, as in the case of qubits. For systems with unbounded spectra, such as harmonic oscillators, $\bar{e}_{\mathrm{max}}$ does not exist, and the work fluctuations can, in principle, grow without limit.

Information-geometric considerations based on the quantum Fisher information (QFI) provide a complementary lower bound of the fluctuations. The QFI for a pure state $|\psi(\lambda)\rangle$ about the parameter $\lambda$ is given by $\mathcal{I}(\lambda) = 4\left[ \langle \partial_\lambda \psi(\lambda) | \partial_\lambda \psi(\lambda)\rangle - |\langle \partial_\lambda \psi(\lambda) | \psi(\lambda)\rangle|^2 \right]$ \cite{Paris_QFI}. Applied to the ground state $|0({\lambda})\ra$, this quantity can be used to prove a universal lower bound on work fluctuations through the inequality 
\begin{equation}
\frac{\sigma^2(\lambda)}{[W(\lambda)]^2}\geq\frac{\sigma^2(\lambda)}{[Q(\lambda)]^2} \geq \frac{2}{\Sigma_Q}, \ \Sigma_Q = \frac{2\mathcal{I}(\lambda)}{\left( \frac{d}{d\lambda} \ln \Delta'(\lambda) \right)^2},
\end{equation}
as derived in the Supplementary Material~\cite{SM}. This bound has the characteristic structure of a thermodynamic uncertainty relation, linking the work to its variance. Here $\Sigma_Q$ plays the role of an effective entropy production associated with heat extraction. In the absence of thermal reservoirs with $T>0$, this bound must be viewed as an extension of thermodynamic uncertainty relations \cite{horowitz2020thermodynamic, barato2015thermodynamic, pietzonka2018universal,cangemi2020violation,miller2021thermodynamic,guarnieri2019thermodynamics} to quantum vacuum measurement engines.  The bound reveals a clear decomposition into an information-theoretic contribution encoded in the QFI and a geometric contribution determined by the variation of the QVBF.

Eq.~\eqref{eq:fluctuation_bounds} can be used to obtain the relation $\text{Var}[\lambda]\geq1/\mathcal{I}(\lambda)$ derived in the Supplementary Material~\cite{SM}.
We note a similarity to the Cram\'er-Rao bound of statistical estimation theory \cite{cramer1999mathematical} and its quantum generalization \cite{braunstein1994statistical}. In that context, the bound on statistical uncertainty can be seen as tied to the sensitivity of the quantum state with respect to that parameter; here thermodynamic twist is that the bound is expressed in terms of work fluctuations, rather than the statistical variance. Nevertheless, by interpreting the work as a parameter-dependent observable, these fluctuations determine the uncertainty of an estimator for $\lambda$.

\textit{Constraints on Engine Performance.}--- We now exploit the qualitative features of ground state energy and expected properties of thermodynamic metrics to derive model-independent bounds on the engine’s performance. 
Let us assume that QVBF is a well-behaved function between $0$ and the value of $\lambda$ at which the engine is operated.
Since the quantum heat satisfies $Q(\lambda) = \lambda \Delta'(\lambda) \geq 0$, 
we have $\Delta'(\lambda)>0$ for $\lambda>0$ and $\Delta'(\lambda)<0$ for $\lambda<0$. Thus, $\Delta(\lambda)$ increases monotonically away from the origin for $\lambda>0$ and decreases monotonically for $\lambda<0$. Because $\Delta(\lambda)$ is positive for $\lambda$ near (but not equal to) $0$, these monotonicity properties imply that $\Delta(\lambda)>0$ for all $\lambda\neq 0$.

The condition $W(\lambda)\geq 0$ implies $\lambda\Delta'(\lambda)\ge \Delta(\lambda)$, constraining the behavior of both the QVBF and $E_0(\lambda)$. This illustrates how fundamental bounds on engine performance restrict the structure of the ground-state energy.

In the weak coupling regime, $\lambda\rightarrow 0$, perturbation theory yields $\Delta(\lambda)\propto \lambda^2$ (see Supplemental Material for details), and we find to leading order in $\lambda$
\begin{eqnarray}
W(\lambda) &=& \frac{\lambda^2 }{2}\Delta''(0), \quad \eta(\lambda) = \frac{1}{2} + \frac{\lambda}{12} \frac{\Delta'''(0)}{\Delta''(0)}, \\ 
\sigma^2(\lambda) &=& \lambda^2 \la 0(0)|H_\text{int}^2|0(0)\ra,    
\end{eqnarray}
indicating that both work and its fluctuations vanish at zero coupling, while the efficiency is universally 1/2 for all quantum vacuum fluctuation engines.

In the strong coupling regime, $\lambda\rightarrow\infty$, the $H_\text{loc}$ acts as a perturbation to $\lambda H_\text{int}$. Provided that the spectrum of the off-diagonal Hamiltonian is bounded, and that possible degeneracies of the ground state of $\lambda H_\text{int}$ are lifted by $H_\text{loc}$, we find that in the large $\lambda$ limit, 
$\Delta(\lambda)\xrightarrow{\lambda\rightarrow \infty}a\lambda+b$, where $a, b$ are real constants. 
Using this asymptotic form in Eq.~\eqref{eq:W_eta_lambda} yields
\begin{equation} \label{eq:high_lambda_limit}
    W(\lambda)\xrightarrow{\lambda\rightarrow \infty}-b, \quad \eta(\lambda)\xrightarrow{\lambda\rightarrow \infty} 0.
\end{equation}
Thus, a quantum measurement engine with a bounded spectrum of $H_\text{int}$ saturates in work output and becomes asymptotically inefficient at strong coupling, a universal feature emerging from the geometry of $\Delta(\lambda)$, that is absent in systems with an unbounded spectrum of $H_\text{int}$.  More generally, we note that the engine can be engineered to near perfect efficiency by ground state engineering. 
If for a given, arbitrary $\lambda$ outside the perturbation theory regime QVBF exhibits locally polynomial scaling, $\Delta \sim c_1 \lambda^p$, then the work scales polynomially $W \sim c_1 (p-1) \lambda^p$, and the efficiency is $\eta = 1-1/p$.  Consequently, for large order $p$ the efficiency approaches 1.  If we have locally exponential scaling, $\Delta \sim c_2 \exp(\beta \lambda)$, then the work scales exponentially, $W \sim c_2 (\lambda \beta -1)\exp(\beta \lambda)$, while the efficiency is $\eta = 1-1/(\beta \lambda)$.  The efficiency approaches 1 for large $\beta$ or $\lambda$.  We conclude that the system should be engineered such that the QVBF increases as rapidly as possible as a function of $\lambda$ to have efficient engines, as implied also by Eq.~\eqref{eq:W_eta_lambda}.

For any $\lambda\neq0$, the extracted work necessarily increases with $|\lambda|$. This follows from the relation
\begin{equation}
    W'(\lambda) = \lambda\Delta''(\lambda),
\end{equation}
together with the result, shown in the Supplemental Material~\cite{SM}, that $\Delta''(\lambda)>0$ for 
$\lambda\neq 0$. This relation provides a direct physical interpretation of the curvature of the QVBF as the rate at which work changes with the coupling strength.

\begin{figure}[t]
    \centering
    \includegraphics[width=1.\linewidth]{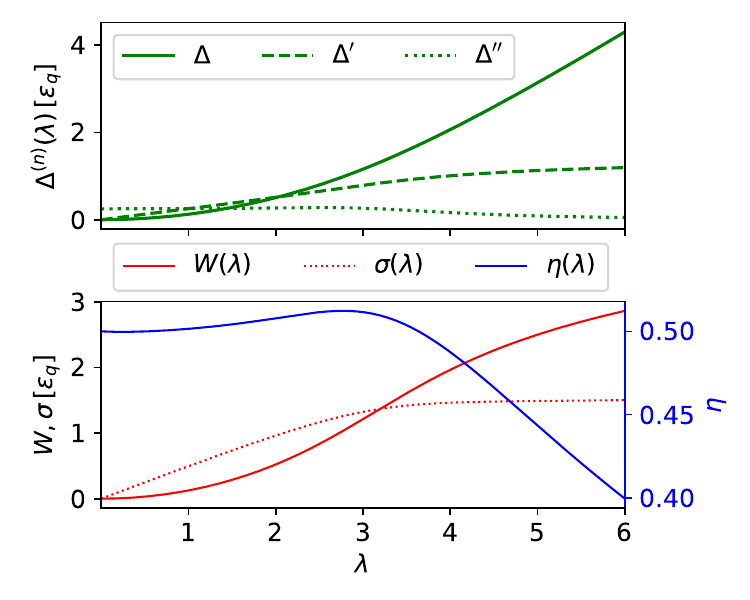}
    \caption{Results for a 10-qubit quantum vacuum measurement engine with randomly chosen couplings, computed using only the quantum vacuum bending function $\Delta(\lambda)$.
    Top panel: the $\Delta(\lambda)$ and its first and second derivatives.
    Bottom panel: work $W$ and its fluctuations $\sigma$ with the corresponding values given on the left y-axis, efficiency $\eta$ with its corresponding values given on the right y-axis. 
    $\Delta(\lambda)$, its derivatives, $W(\lambda)$, and $\sigma(\lambda)$ are plotted in units of the qubit energy gap $\varepsilon_q$.
    The coupling values used to generate the data are provided in the Supplementary Material~\cite{SM}.}
    \label{fig:results_10_qubits}
\end{figure}

Taken together, all considered coupling regimes show that the thermodynamic performance of a quantum measurement engine is fully constrained by the slope and curvature of $\Delta(\lambda)$. These conclusions are independent of the microscopic details of the system and therefore apply generically to quantum measurement engines across different physical platforms.

\textit{Qubit-based engine example.}---As a concrete illustration of the general framework, we now analyze a quantum measurement engine implemented using a chain of $N$ coupled qubits. The Hamiltonian is given by 
\begin{equation} \label{eq:10_qubits}
H_q(\lambda)=\frac{\omega}{2}\sum_{j=1}^N\sigma_{j}^z + \frac{\lambda}{2}\sum_{j=1}^N\sum_{k=j+1}^N g_{jk} \sigma_j^x  \sigma_k^{x},
\end{equation}
where $\sigma_{j}^{z(x)}$ are the Pauli matrices for qubit $j$, $\varepsilon_q$ is the qubit energy gap and $g_{jk}$ are the coupling strengths. This model provides a setting in which the QVBF $\Delta(\lambda)$ acquires a nontrivial shape and allows us to validate the geometric relations derived in the previous sections.

\begin{figure}
    \centering
    \includegraphics[width=1.\linewidth]{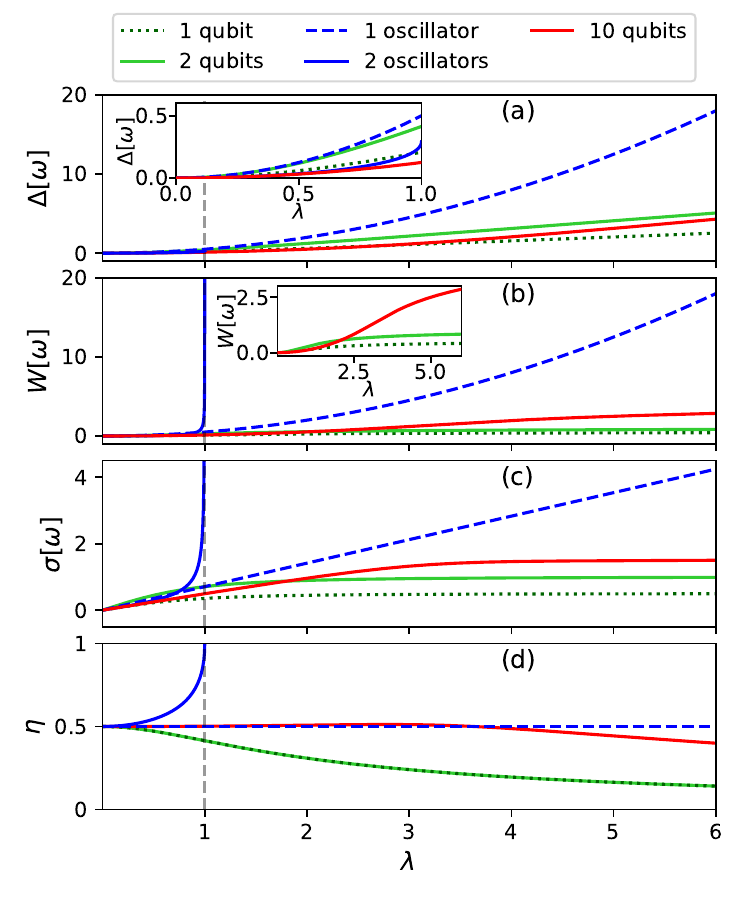}
    \caption{Comparison of five types of measurement engines described in Tab.~\ref{tab:descriptions}: single qubit, coupled qubits, single harmonic oscillator, coupled harmonic oscillators, and a $10$-qubit engine with randomly selected couplings. The red curves correspond to the $10$-qubit results shown in Fig.~\ref{fig:multiple_engines}. Panel (a) shows the entanglement gap $\Delta(\lambda)$; the inset displays the same quantity over a smaller range of $\lambda$. Panel (b) presents the extracted work, with the inset restricted to qubit-based engines. Panel (c) shows the work fluctuations, and panel (d) displays the engine efficiency. The energy unit $\omega$ denotes the qubit energy gap for qubit-based engines and the level spacing for oscillator-based engines.}
    \label{fig:multiple_engines}
\end{figure}

Figure~\ref{fig:results_10_qubits} presents numerical results for a quantum measurement engine constructed from $N=10$ qubits with randomly chosen coupling strengths provided in Supplementary Material~\cite{SM}. We numerically verified that the ground state remains nondegenerate over the entire range of $\lambda$ shown, ensuring the validity of the theoretical framework developed above. This plot confirms the features predicted in our work. Near origin $\lambda=0$ the work increases quadratically, and efficiency starts at $\eta(0)=\tfrac{1}{2}$. Work increases monotonically for all considered $\lambda$. Fluctuations $\sigma$ and efficiency $\eta$ have a linear change away from the origin. For large $\lambda$, the work saturates to a finite value and efficiency drops to $0$. This example illustrates that the full thermodynamic performance of the engine is determined by the geometry of the QVBF.

Figure~\ref{fig:multiple_engines} and Table~\ref{tab:descriptions} compares quantum measurement engines implemented with qubits and harmonic oscillators. While qubit-based engines exhibit saturation of the extracted work at strong coupling, oscillator-based engines do not. This distinction originates from the spectral properties of the interaction Hamiltonian $H_\text{int}$: for qubits, $H_\text{int}$ has a bounded spectrum and a well-defined ground state, whereas for oscillators it is unbounded. The boundedness of the $H_\text{int}$ spectrum enforces saturation of the work output, as discussed in Eq.~\eqref{eq:high_lambda_limit}. Consistently, work fluctuations remain bounded for qubit-based engines but grow without bound for oscillator-based engines, in agreement with Eq.~\eqref{eq:fluctuation_bounds}. For the two-oscillator engine, the Hamiltonian ceases to have a ground state when $|\lambda|\ge \omega^2$, rendering the engine non-operational in this regime. Although the behavior near this point suggests divergent work extraction, achieving it would require supplying infinite energy to the system by the measurement apparatus. 
The engines based on two coupled oscillators and the oscillator chain considered in Supplementary Material exhibit near-unit efficiency when the fluctuations diverge, analogous to the results presented in Ref.~\cite{pietzonka2018universal} for engines driven by temperature gradient between two baths. All plotted results in Fig.~\ref{fig:multiple_engines} are derived in Supplementary Material~\cite{SM} except red curves which were calculated numerically. There, we also exactly solve the case of thermodynamic systems, where we derive QVBF for the continuum fermonic limit of the transverse field Ising model, as well as for of the continuum bosonic limit of an oscillator chain, where QVBF can be expressed in terms of the complete Elliptic integral of the second kind.  This function exhibits logarithmic singularities, indicating phase transitions and anomalous engine characteristics at those points.

\renewcommand{\arraystretch}{1.2}
\begin{table}[t]
\centering
\begin{tabular}{cc}
\hline
\textbf{Description} & \textbf{Hamiltonian} \\ \hline
1 qubit   &  $\omega \sigma^z / 2 + \lambda \sigma^x/ 2$ \\ \hline
2 qubits  & $\omega \sigma_1^z / 2 + \omega \sigma_2^z / 2 + \lambda \sigma_1^x \sigma_2^x$  \\ \hline
10 qubits & Eq.~\eqref{eq:10_qubits}       \\ \hline
1 oscillator  & $p^2/2 + \omega^2 x^2/2 + \lambda x$        \\ \hline
2 oscillators & $\quad p_1^2/2 + \omega^2 x_1^2/2+ p_2^2/2 + \omega^2 x_2^2/2 + \lambda x_1 x_2$    \\ \hline
\end{tabular}
\caption{Hamiltonians used to produce Fig.~\ref{fig:multiple_engines}. We use units $\hbar = 1$.}
\label{tab:descriptions}
\end{table}

\textit{Conclusion.}---We have developed a general geometric theory of quantum vacuum measurement engines, extending and unifying the framework introduced in Ref.~\cite{jussiau2023many}. Our central result is that the quantum vacuum bending function $\Delta(\lambda)$—encoding the ground-state energy landscape as a function of coupling strength—fully determines the engine’s thermodynamic behavior. In particular, the average work, quantum heat, and efficiency depend solely on $\Delta(\lambda)$ and its first derivative, while work fluctuations are governed by its curvature weighted by an effective excitation energy. This establishes the ground-state energy landscape as the fundamental object controlling engine performance, independently of the microscopic details of the working medium.

We have demonstrated how the features characterizing the ground state energy landscape can be translated into performance metrics of the engine. Constraints on work fluctuations take the form of an extension of a thermodynamic uncertainty relation adapted to quantum vacuum measurement engines, where the role of entropy production is played by a combination of quantum Fisher information and the geometry of $\Delta(\lambda)$. Therefore, the fluctuations can be used to probe information-theoretic metrics. Our results suggest a novel way of designing quantum thermal machines, where manipulating the ground state energy landscape can enhance their performance. The quantum vacuum bending function emerges as a physically relevant quantity beyond engine operation, linking quantum thermodynamics, metrology, and many-body correlations within a single geometric framework.

{\it Data availability statement.---}Source code and data used to produce the plots in this manuscript are available from the corresponding author upon reasonable request. 

{\it Acknowledgements.---}This work was supported by the John Templeton Foundation Grant ID 63209. We thank Ketevan Kotorashvili for her assistance with figure design, and Abhaya S. Hegde and Taylor Lee Patti for helpful discussions.

\bibliography{references}

\clearpage
\pagestyle{empty}
\onecolumngrid
\vspace*{10pt}
\renewcommand{\theequation}{S\arabic{equation}}
\setcounter{equation}{0}
\begin{bibunit}[apsrev4-2]
\begin{center}
	\large \textbf{Supplemental Material for ``Universal Characterization of Quantum Vacuum Measurement Engines"}
\end{center}

\section{Derivation of fluctuations}

In this section, we derive the result
\begin{equation} \label{eq:fluctuations}
    \sigma^2(\lambda) = \frac{1}{2}\lambda^2 \Delta''(\lambda) \bar{e}(\lambda)
\end{equation}
describing the fluctuations of the engine. The calculations are divided into parts and presented in the following subsections.

\subsection{Derivative of the ground state of the total Hamiltonian}

We aim to derive the expression for $|0'(\lambda)\rangle$. To keep the notation concise, we will omit the dependence on $\lambda$ and write it as $|0'\rangle$ and similar for the other $\lambda$-dependent quantities. First, consider the norm
\begin{equation}
    \langle 0|0\rangle =1,
\end{equation}
and differentiate it over $\lambda$ to get
\begin{equation}
    2\Re\left\{ \langle 0|0'\rangle \right\}=0.
\end{equation}
Here the prime indicates the derivative with respect to $\lambda$; the result implies that the overlap of the ground state and its derivative must be purely imaginary
\begin{equation}
    \langle 0|0'\rangle = i f(\lambda),
\end{equation}
where $f(\lambda)$ is a real function. We will use this property in future calculations. 

Now consider the eigenproblem
\begin{equation}
    H|0\rangle=E_0|0\rangle,
\end{equation}
and take the derivative of both sides over $\lambda$ to get
\begin{equation} \label{eq:eigenproblem_derivative}
    H'|0\rangle+H|0'\rangle = E_0'|0\rangle + E_0|0'\rangle.
\end{equation}
Applying the eigenstate $\bra{n}$ of the total Hamiltonian $H(\lambda)$ from the left results in 
\begin{equation}
    \langle n|H'|0\rangle+E_n\langle n|0'\rangle = E_0\langle n|0'\rangle,
\end{equation}
and manipulating this expression gives
\begin{equation}
    \langle n|0'\rangle = - \frac{\langle n|H'|0\rangle}{E_n-E_0} = - \frac{\langle n|H_\text{int}|0\rangle}{E_n-E_0}.
\end{equation}

Let us come back to the ground state derivative
\begin{equation} \label{eq:d0_dlambda}
\begin{aligned}
    |0'\rangle & = \sum_n |n\rangle\langle n|0'\rangle \\
    & = |0\rangle\langle0|0'\rangle+\sum_{n>0} |n\rangle\langle n|0'\rangle \\
    & = if(\lambda)|0 \rangle - \sum_{n>0} |n\rangle \frac{\langle n|H_\text{int}|0\rangle}{E_n-E_0}.
\end{aligned}
\end{equation}

Using this expression to evaluate $\langle 0'|0'\rangle$ results in
\begin{equation} \label{eq:d0_norm}
    \langle 0'|0'\rangle = f(\lambda)^2+\sum_{n>0}\frac{|\langle n|H_\text{int}|0\rangle|^2}{(E_n - E_0)^2}.
\end{equation}

\subsection{Second derivative of the ground state energy}

We begin with the result obtained through Hellmann-Feynman theorem
\begin{equation}
    E_0' = \langle0|H_\text{int}|0\rangle,
\end{equation}
and differentiate it over $\lambda$. It gives
\begin{equation}
    E_0'' = (\langle0|H_\text{int}|0\rangle)'=2\Re\{ \langle0|H_\text{int}|0'\rangle \}.
\end{equation}

Using \eqref{eq:d0_dlambda} results in
\begin{equation}\label{eq:d2e0_dlambda2}
    E_0''=-2\sum_{n>0} \frac{|\langle n|H_\text{int}|0\rangle|^2}{E_n-E_0}.
\end{equation}
This expression must be non-positive, implying that $\Delta''=-E_0''$ must be non-negative. Equality is reached only for $\lambda=0$. $\Delta''>0$ implies that $\Delta$ is a concave up function which grows indefinitely as a function of $\lambda$.

\subsection{Variance of the interaction Hamiltonian}

Consider the expected value of the square of the interaction Hamiltonian evaluated in the ground state
\begin{equation} \label{eq:second_moment}
    \expval{H_\text{int}^2} = \sum_n \langle 0 | H_\text{int} |n\rangle \langle n | H_\text{int} |0\rangle = \langle 0 | H_\text{int} |0\rangle^2  + \sum_{n>0} \langle 0 | H_\text{int} |n\rangle \langle n | H_\text{int} |0\rangle,
\end{equation}
which implies
\begin{equation} \label{eq:h_int_fluctuations}
    \expval{H_\text{int}^2} - \expval{H_\text{int}}^2 =  \sum_{n>0}  |\langle n | H_\text{int} |0\rangle|^2,
\end{equation}
and results in the variance we seek. 

\subsection{Variance of the work output}\label{sec:variance}

We begin with the expression
\begin{equation}\label{eq:sigma}
    \sigma^2 = \expval{H_{\text{loc}}^2}_{\lambda} - 
              \expval{H_{\text{loc}}}_{\lambda} ^2 = \lambda^2\left( \expval{H_\text{int}^2}_{\lambda} - \expval{H_\text{int}}_{\lambda} ^2 \right),
\end{equation}
which connected with result \eqref{eq:h_int_fluctuations} gives
\begin{equation} \label{eq:sigma_v1}
    \sigma^2 = \lambda^2 \sum_{n>0}  |\langle n | H_\text{int} |0\rangle|^2.
\end{equation}
This expression implies that the fluctuations emerge from the couplings between the ground and excited states through the interaction Hamiltonian. It is the main reason why $\sigma^2$ cannot be expressed solely in terms of the quantum vacuum bending function (QVBF) $\Delta$ and its derivatives-they are dependent on the second moment of $\expval{H_\text{int}^2}$, which couples to the excited states of the total Hamiltonian [see Eq.~\eqref{eq:second_moment}]. This coupling is explicit in $\eqref{eq:sigma_v1}$.

We can re-express the fluctuations to find explicit dependence on QVBF $\Delta$, its derivatives, and corrections introduced by couplings to the excited states. Let us define 
\begin{equation} \label{eq:cn_en}
    c_n \equiv |\langle n | H_\text{int} |0\rangle|, \quad e_n \equiv E_n-E_0.
\end{equation}
Then 
\begin{equation}
    \expval{H_\text{int}^2} - \expval{H_\text{int}}^2 =  \lambda^2 \sum_{n>0} c_n^2,
\end{equation}
\begin{equation}
    \Delta'' = -E_0''=2\sum_{n>0} \frac{c_n^2}{e_n},
\end{equation}
where the last expression was obtained from Eq.~\eqref{eq:d2e0_dlambda2}. 

Let us define $\bar{e}$ as the weighted average of $1/e_n$ weighted by the coefficients $c_n^2$
\begin{equation} \label{eq:1_bar_e}
    \frac{1}{\bar{e}} \equiv \frac{\sum_{n>0} (c_n^2/e_n)}{\sum_{n>0} c_n^2} = \frac{\Delta''/2}{\sigma^2/\lambda^2}.
\end{equation}
The energy $\bar{e}$ is bounded by the maximum and minimum energy difference between the excited state and ground state.
\begin{equation}
    e_1 \leq \bar{e} \leq e_{\text{max}},
\end{equation}
where $e_1 = E_1 - E_0$, $e_{\text{max}} = \max_n(E_n-E_0)$.
From \eqref{eq:1_bar_e} we have
\begin{equation} \label{eq:sigma_lambda}
    \sigma^2 = \frac{1}{2}\lambda^2 \Delta'' \bar{e}
\end{equation}
which is the expression provided in the manuscript. 

\section{Comments on perturbation theory}

\subsection{Small coupling regime}

Let us consider the behavior of QVBF in small $|\lambda|$ regime. We assume that the ground state of the total Hamiltonian is non-degenererate within the perturbation theory regime.
The important ingredient of our calculations is the requirement
\begin{equation} \label{eq:condition}
    \langle n(\lambda=0)|H_\text{int}|n(\lambda=0)\rangle=0,
\end{equation}
which was assumed for the engine. We can approach the small $|\lambda|$ regime in two ways, either through series expansion or through perturbation theory, where both approaches are expected to return the same result. We emphasize that the results demonstrated in the main manuscript assume non-degenerate ground state $|0(\lambda)\rangle$, which justifies the use of non-degenerate perturbation theory in computing the ground state energy. 

Consider expanding QVBF in Maclaurin series
\begin{equation}
    \Delta(\lambda)= \Delta(0)+\Delta'(0)\lambda + \frac{1}{2}\Delta''(0)\lambda^2 + \mathcal{O}(\lambda^3),
\end{equation}
where we will neglect $\mathcal{O}(\lambda^3)$ as the previous terms are expected to be dominant. The definition of QVBF implies $\Delta(0)=0$. The first derivative can be computed by employing Hellman-Feynman theorem 
\begin{equation}
    \Delta'(0)=-E_0'(0) = -\langle0(\lambda=0)|H_\text{int}|0(\lambda=0)\rangle =0,
\end{equation}
where in the final line we used Eq~\eqref{eq:condition}. The second derivative can be calculated by noticing $\Delta''(\lambda)=-E_0''(\lambda)$ and using Eq.~\eqref{eq:d2e0_dlambda2}. This implies that near the origin QVBI behaves like a function quadratic in $\lambda$
\begin{equation}
    \Delta(\lambda) \xrightarrow{\lambda\rightarrow 0} \frac{1}{2}\Delta''(0)\lambda^2.
\end{equation}
The same conclusions can be obtained through non-degenerate perturbation theory, where $\lambda H_\text{int}$ is a perturbation to $H_\text{loc}$. The ground state can be expanded as
\begin{equation}
    E_0(\lambda)=E_0(0) + \lambda E_0^{(1)} + \lambda^2 E_0^{(2)} + \mathcal{O}(\lambda^3),
\end{equation}
where $E_0^{(n)}$ is the $n$-th order correction to the ground state energy.
Using the definition $\Delta(\lambda) = E_0(0)-E_0(\lambda)$ yields
\begin{equation}
    \Delta(\lambda) =  - \lambda E_0^{(1)} -  \lambda^2 E_0^{(2)} + \mathcal{O}(\lambda^3).
\end{equation}
The first-order correction to the ground state energy $|0(0)\rangle=|0(\lambda=0)\rangle$ is 
\begin{equation}
    E_0^{(1)} = \langle 0(0)|H_\text{int} | 0(0) \rangle = 0,
\end{equation}
which vanishes due to the assumed constraint \eqref{eq:condition}. The second order correction is 
\begin{equation}
    E_0^{(2)} = \sum_{n>0} \frac{|\langle n(0)|H_\text{int}|0(0)\rangle|^2}{E_0(0) - E_n(0)},
\end{equation}
resulting in 
\begin{equation}
    \Delta(\lambda)\xrightarrow{\lambda\rightarrow0}  \lambda^2 \sum_{n>0} \frac{|\langle n(0)|H_\text{int}|0(0)\rangle|^2}{E_n(0) - E_0(0)}  = -\frac{1}{2}E_0''(\lambda=0)\lambda^2,    
\end{equation}
which agrees with Eq.~\eqref{eq:d2e0_dlambda2} evaluated at $\lambda=0$. 

Expressed in terms of a Taylor series in $\lambda$, the small $\lambda$ behavior of work $W$ can be seen because $\Delta$ has its first non-vanishing order at $\lambda^2$, thus 
\be
W = \lambda \Delta'(\lambda) - \Delta(\lambda) =\frac{1}{2} \Delta''(0) \lambda^2 + {\cal O}(\lambda^3).
\ee
The efficiency can be expanded to first order in $\lambda$ as
\begin{equation}\label{eq:eta_small_lambda}
\begin{aligned}
\eta(\lambda) &= 1 - \frac{ \Delta(\lambda)}{\lambda \Delta'(\lambda)} \\
&=1 - \frac{(1/2) \Delta''(0) \lambda^2+ (1/6) \Delta'''(0)\lambda^3 }{\Delta''(0) \lambda^2 + (1/2) \Delta'''(0) \lambda^3} \nonumber \\
&=
\frac{1}{2} + \frac{1}{12} \frac{\lambda \Delta'''(0)}{\Delta''(0)} +  {\cal O}(\lambda^2).
\end{aligned}
\end{equation} 
Here, the linear term may vanish if have $\Delta'''(0)=0.$

The fluctuations of work, (\ref{eq:sigma_v1}) begin at order $\lambda^2$, so this order of perturbation theory can be obtained by setting $\lambda=0$ in the other factors, yielding
\begin{eqnarray} \label{eq:sigma_pert}
    \sigma^2(\lambda) &=& \lambda^2 \sum_{n>0}  |\langle n(0) | H_\text{int} |0(0)\rangle|^2\\
&=& \lambda^2 \sum_{n>0} \la 0(0)| H_\text{int} | n(0) \ra \la n(0) | H_\text{int} |0(0)\rangle\\
&=&     \lambda^2 \la 0(0)| H_\text{int}^2 |0(0)\rangle + {\cal O}(\lambda^3),
\end{eqnarray}
where we used the completeness of the zero coupling states and the fact that we assume $\la 0(0) | H_\text{int} | 0(0) \ra =0$.

\subsection{High coupling regime}

Let us now consider the regime of large coupling strength $|\lambda|$, in which the local Hamiltonian $H_\text{loc}$ can be treated as a perturbation to the interaction Hamiltonian $H_\text{int}$. The purpose of this subsection is to qualitatively characterize the behavior of QVBF in this regime.

We assume that $H_\text{int}$ possesses a ground state, which implies that $\lambda H_\text{int}$ also has a ground state for $\lambda>0$. Even if the ground state of $\lambda H_\text{int}$ is degenerate, we assume that this degeneracy is lifted by the perturbation $H_{\text{loc}}$, resulting in a non-degenerate ground state of the full Hamiltonian.

We denote by $|0_{\text{int}}\rangle$ the ground state of $\lambda H_\text{int}$, or, in the degenerate case, the particular superposition of degenerate ground states that minimizes the lowest-order energy correction due to $H_\text{loc}$. Since the total Hamiltonian is assumed to have a non-degenerate ground state, this superposition is taken to be unique.

By definition, $|0_{\text{int}}\rangle$ satisfies
\begin{equation}
    H_\text{int}|0_\text{int}\rangle = E_0^\text{int} |0_\text{int}\rangle,
\end{equation}
implying
$\lambda H_\text{int}|0_\text{int}\rangle = \lambda E_0^\text{int} |0_\text{int}\rangle$. 

We anticipate that the lowest-order correction to the ground-state energy arising from $H_\text{loc}$ scales as $\lambda^{-p}$ with $p \ge 0$, where the case $p=0$ corresponds to a non-vanishing first-order correction. Accordingly, the ground-state energy at large $|\lambda|$ is expected to behave as
\begin{equation} \label{eq:ansatz_e0_lambda}
    E_0(\lambda) \approx \lambda E_0^\text{int} + \lambda^{-p} E_p, 
\end{equation}
where $E_p$ is a correction term independent of $\lambda$. The precise form of this contribution does not affect the leading-order conclusions. QVBF is then
\begin{equation}
    \Delta(\lambda) = E_0(0) - E_0(\lambda) \approx  - \lambda E_0^\text{int} + E_0(0)- \lambda^{-p} E_p.
\end{equation}
The term $E_0(0)$ cannot be inferred from the large-$|\lambda|$ expansion, as it is defined in a different coupling regime. When the first-order degenerate correction is active, the remaining terms beyond the linear contribution scale as constants.

Since both $E_0(0)$ and the possible correction in $p=0$ case are independent of $\lambda$ at leading order, we adopt the simplified ansatz
\begin{equation}
    \Delta(\lambda) \approx - \lambda E_0^\text{int} + C,
\end{equation}
where $C$ is a constant. The corresponding derivatives are
\begin{eqnarray}
    \Delta'(\lambda) \approx - E_0^\text{int}, \quad \Delta''(\lambda) \approx 0. 
\end{eqnarray}
Using the relations $Q(\lambda)=\lambda\Delta'(\lambda)$ and $W(\lambda)=\lambda\Delta'(\lambda)-\Delta(\lambda)$, we obtain the expected asymptotic scalings in the high-$|\lambda|$ regime:
\begin{equation}
\begin{aligned}
    Q(\lambda) & \propto \lambda, \\
    W(\lambda) & \propto \text{const}, \\
    W'(\lambda) & \propto 0, \\
    \eta(\lambda) & \propto 1/|\lambda|\xrightarrow{|\lambda|\rightarrow\infty} 0.
\end{aligned}
\end{equation}
These results imply that the work output saturates at large coupling, while the efficiency vanishes asymptotically. We emphasize that our considerations in the high-$|\lambda|$ regime apply only if the ground state of $\lambda H_\text{int}$ exists; these results are not applicable otherwise.

\section{Work and engine's operational regimes}

Consider the behavior of $W(\lambda)$ for $\lambda\geq0$ and bounded spectrum of $H_\text{int}$: it begins at $0$, starts with quadratic growth, and saturates for large $\lambda$. It implies that there must be a critical point $\lambda_c$ below which it is beneficial to increase coupling as it results in super-linear growth of $W(\lambda)$. Above this point, sublinear growth is expected and increased coupling does not lead to significant increases in $W(\lambda)$. We will call these regimes \textbf{convex-growth regime}, and \textbf{concave-saturation regime} respectively.

Recall that $W'(\lambda) = \lambda\Delta''(\lambda)$, resulting in
\begin{equation}
    W''(\lambda) = \Delta''(\lambda) + \lambda\Delta'''(\lambda).
\end{equation}
The boundary between the regimes can be found from
\begin{equation}
    W''(\lambda_c) = \Delta''(\lambda_c) + \lambda_c\Delta'''(\lambda_c) =0,
\end{equation}
and solving this equation results in the critical point $\lambda_c$. 

A similar argument can be made for negative $\lambda$. However, the critical points on both negative and positive branches will not be located at symmetric locations, and their symmetry is expected only when $\Delta(\lambda)=\alpha\Delta(-\lambda)$, where $\alpha$ is a positive constant independent of $\lambda$. However, this relation is in general not satisfied.

\section{Integral expressions for work and quantum heat}

Suppose that $\Delta(\lambda)$ is a well-behaved function between $0$ and $\lambda$. The expression for work provided in the manuscript is
\begin{equation} \label{eq:work}
    W(\lambda) = \lambda \Delta'(\lambda) - \Delta(\lambda).  
\end{equation}
This can be recovered from the integral
\begin{equation} \label{eq:work_int}
    W(\lambda) = \int_0^x x \Delta''(x) dx,
\end{equation}
where integration by parts returns \eqref{eq:work}. 

Quantum heat satisfies $Q(\lambda)=W(\lambda)+\Delta(\lambda)$, and we can use fundamental theorem of calculus to write 
\begin{equation}
    \Delta(\lambda) = \int_0^{\lambda} \Delta'(x) dx.
\end{equation}

Adding this to \eqref{eq:work_int} yields
\begin{equation}\label{eq:heat_int}
    Q(\lambda) = \int_0^\lambda \left[ x \Delta''(x) + \Delta'(x) \right]dx.
\end{equation}

Eqs.~\eqref{eq:work_int} and \eqref{eq:heat_int} represent an alternative, integral formulation of the results derived in the main paper. 

\section{Bounds on fluctuations}

The quantum Fisher information (QFI) for a pure state $|0(\lambda)\rangle$ concerning the parameter $\lambda$ is given by
\begin{equation}
    \mathcal{I}(\lambda) = 4\left[ \langle 0'(\lambda) | 0'(\lambda)\rangle - |\langle 0'(\lambda) | 0(\lambda)\rangle|^2 \right].
\end{equation}
Using \eqref{eq:cn_en} and \eqref{eq:d0_dlambda}, this expression can be written as
\begin{equation} \label{eq:qfi_v2}
    \mathcal{I}(\lambda) = 4 \sum_{n>0} \frac{c_n^2}{e_n^2}.
\end{equation}
Applying the Cauchy–Schwarz inequality yields
\begin{equation} \label{eq:cauchy_schwartz}
    \left( \sum_{n>0} \frac{c_n}{e_n} c_n \right)^2 \leq \left( \sum_{n>0} \frac{c_n^2}{e_n^2}\right)\left( \sum_{n>0} c_n^2\right).
\end{equation}
Eq.~\eqref{eq:qfi_v2} together with the results derived in Sec.~\ref{sec:variance}, implies
\begin{equation}
\begin{aligned}
    & \left( \sum_{n>0} \frac{c_n}{e_n} c_n \right) = \frac{\Delta''(\lambda)}{2}, \\
    & \left( \sum_{n>0} \frac{c_n^2}{e_n^2}\right) = \frac{\mathcal{I}(\lambda)}{4}, \\
    & \left( \sum_{n>0} c_n^2\right) = \frac{\sigma^2(\lambda)}{\lambda^2}.
\end{aligned}
\end{equation}
Substituting these relations into the Cauchy–Schwarz inequality gives
\begin{equation}
    \sigma^2(\lambda) \geq \frac{\lambda^2\left[\Delta''(\lambda)\right]^2}{\mathcal{I}(\lambda)} = \frac{[W'(\lambda)]^2}{\mathcal{I}(\lambda)},
\end{equation}
demonstrating that the curvature of QVBF, together with the inverse of the QFI, provides a lower bound on the fluctuations.

We now consider fluctuations relative to the quantum heat, using $Q(\lambda)=\lambda\Delta'(\lambda)$. Dividing the above inequality by $Q^2(\lambda)$ yields
\begin{equation}
    \frac{\sigma^2(\lambda)}{[Q(\lambda)]^2} \geq \frac{2}{\Sigma_Q}, \quad \Sigma_Q = 2 \mathcal{I}(\lambda)\left( \frac{\Delta'(\lambda)}{\Delta''(\lambda)} \right)^2,
\end{equation}
This expression has the same structure as thermodynamic uncertainty relations (TURs), where $\Sigma_Q$ plays the role of an effective entropy production associated with the quantum heat current. Finally, $\Sigma_Q$ can be expressed in an equivalent form as
\begin{equation}
    \Sigma_Q = \frac{2\mathcal{I}(\lambda)}{\left( \frac{d}{d\lambda} \ln \Delta'(\lambda) \right)^2}.
\end{equation}

This representation makes explicit that $\Sigma_Q$ consists of two contributions: an information-theoretic component encoded in the QFI, and a geometric component determined by the derivative of QVBF.

Finally, since $W(\lambda) = \eta(\lambda)Q(\lambda)$, and $0\leq\eta(\lambda)\leq1$, we must have
\begin{equation}
    \frac{\sigma^2(\lambda)}{[W(\lambda)]^2}\geq\frac{\sigma^2(\lambda)}{[Q(\lambda)]^2} \geq \frac{2}{\Sigma_Q}. \label{ineq}
\end{equation}

We can view work $W(\lambda)$ as a parameter-dependent observable, where its fluctuations determine the uncertainty of an estimator for $\lambda$.
To make the notion more precise, we make error propagation, $\delta W = W'(\lambda)\delta \lambda$.  Let us consider the variance of $\lambda$ as a statistical uncertainty though the work fluctuations
\begin{equation}
    \sigma^2 = \la \delta W^2\ra = (W'(\lambda))^2 \la \delta \lambda^2\ra.
\end{equation}
From the previous result (\ref{eq:work}) we have $W'(\lambda) = \lambda \Delta''$.  Using the inequality (\ref{ineq}), we have
\be
\text{Var}[\lambda] = \la \delta \lambda^2\ra = \frac{\sigma^2}{Q^2} \frac{Q^2}{\lambda^2 (\Delta'')^2} \ge 1/\cal I,
\ee
where we have used $Q = \lambda \Delta'$.
This is the same as the quantum Cram\'er-Rao bound.

\section{Examples}

The following sections contain the analytic and numerical computations describing the physical systems selected to validate the theory presented in the manuscript.

\subsection{Coupled qubits}
Let us consider the engine formed by $N=2$ qubits. The Hamiltonian is
\begin{equation}
    H(\lambda) = \frac{\omega}{2}\sigma_z^{(1)} + \frac{\omega}{2}\sigma_z^{(2)} + \lambda \sigma_x^{(1)} \sigma_x^{(2)}.
\end{equation}
We have
\begin{equation}
    H_{\text{loc}} = \frac{\omega}{2}\sigma_z^{(1)} + \frac{\omega}{2}\sigma_z^{(2)}, \quad H_\text{int} = \sigma_x^{(1)} \sigma_x^{(2)}
\end{equation}
It represents one of the simplest non-trivial example of an interacting system we can use to test the theory developed in the main body of the manuscript. This realization of vacuum measurement engine was considered by Juissau \textit{et al.}, it defines the local entanglement gap which is equivalent to QVBF in case of many-body quantum vacuum measurement engines.
The equations below represents their results adapted to the notation including the interaction tuning parameter $\lambda$. The energy eigenstates are
\begin{equation}
\begin{aligned}
    |\phi^+\rangle & = \sin\frac{\phi}{2}|00\rangle + \cos\frac{\phi}{2}|11\rangle, \\
    |\psi^+\rangle & = (|01\rangle+|10\rangle)/\sqrt{2}, \\
    |\psi^-\rangle & = (|01\rangle-|10\rangle)/\sqrt{2}, \\
    |\phi^-\rangle & = \cos\frac{\phi}{2}|00\rangle - \sin\frac{\phi}{2}|11\rangle, 
\end{aligned}
\end{equation}
where
\begin{equation}
    \cos\phi=\frac{\omega}{\sqrt{\omega^2+\lambda^2}}, \quad \tan\phi = \frac{\lambda}{\omega}.
\end{equation}
The corresponding eigenenergies are
\begin{equation}
    E_\psi^{\pm}=\pm\lambda, \quad E_\phi^\pm = \pm \sqrt{\omega^2+\lambda^2}.
\end{equation}
The ground state is 
\begin{equation}
    E_0(\lambda) = E_\phi^-.
\end{equation}
The difference between the ground state and the first excited state is 
\begin{equation}
    E_\psi^- - E_\phi^- = - \lambda + \sqrt{\omega^2+\lambda^2},
\end{equation}
and its always positive by definition, implying that the theory developed in this paper can be applied. QVBF is 
\begin{equation}
    \Delta(\lambda) = E_0(0)-E_0(\lambda) = - \omega + \sqrt{\omega^2 + \lambda^2},
\end{equation}
and its positive for $\lambda\neq0$. Adapting the expressions for engine performance metrics derived by Juissau \textit{et al.} results in
\begin{equation} \label{eq:w_2qubits}
    W(\lambda) = \omega - \frac{\omega^2}{\sqrt{\omega^2+\lambda^2}},
\end{equation}
\begin{equation}
    Q(\lambda)=\frac{\lambda^2}{\sqrt{\omega^2+\lambda^2}},
\end{equation}
\begin{equation} \label{eq:eta_2qubits}
    \eta(\lambda) = \frac{\omega}{\lambda^2}\left(  \sqrt{\omega^2+\lambda^2} - \omega\right),
\end{equation}
\begin{equation} \label{eq:sigma_2qubits}
    \sigma^2(\lambda)=\frac{\lambda^2\omega^2}{\omega^2+\lambda^2}
\end{equation}
We will now verify if these expressions agree with the theory derived in this manuscript. The derivative of QVBF is
\begin{equation} \label{eq:dDelta_2qubits}
    \Delta'(\lambda)=\frac{\lambda}{\sqrt{\omega^2+\lambda^2}},
\end{equation}
and when multiplied by $\lambda$
\begin{equation}
\lambda\Delta'(\lambda)=\frac{\lambda^2}{\sqrt{\omega^2+\lambda^2}} = Q(\lambda),
\end{equation}
as expected. Subtracting $\Delta(\lambda)$ from this expression yields after some algebra
\begin{equation}
    \lambda\Delta'(\lambda) - \Delta(\lambda) = \omega - \frac{\omega^2}{\sqrt{\omega^2+\lambda^2}} = W(\lambda).
\end{equation}
The efficiency can be obtained through division $\eta(\lambda)-W(\lambda)/Q(\lambda)$; therefore, since work and heat agree with the results derived by Juissau \textit{et al.}, the efficiency must agree as well. The second derivative of QVBF is 
\begin{equation} \label{eq:d2Delta_2qubits}
    \Delta''(\lambda)=\frac{\omega^2}{\left(\omega^2+\lambda^2\right)^{3/2}}.
\end{equation}
Now consider the coupling coefficients $c_n$ defined in Eq.~\eqref{eq:cn_en}. $H_\text{int}$ connects only states with the same parity, therefore the ground state can only couple to the state $|\psi^+\rangle$. Since there is only one active coefficient $c_n$, we have
\begin{equation}
    \bar{e}=E_\phi^+ - E_\phi^- = 2\sqrt{\omega^2 +\lambda^2}.
\end{equation}
Combining this expression with \eqref{eq:sigma_lambda} and \eqref{eq:d2Delta_2qubits} returns \eqref{eq:sigma_2qubits}, in agreement with the previous work. 

Let us now examine the laws describing the connection between thermodynamic properties and the shape of $\Delta(\lambda)$. We claimed that since $Q(\lambda)\geq 0$, we must have $\Delta'(\lambda)>0$ for $\lambda>0$, and $\Delta'(\lambda)<0$ for $\lambda<0$. Eq.~\eqref{eq:dDelta_2qubits} agrees with this prediction. 

Now, let us consider the low $|\lambda|$ regime. Expanding the work in Taylor series yields
\begin{equation}
    W(\lambda) = \frac{\lambda^2}{2\omega} + \mathcal{O}(\lambda^4),
\end{equation}
i.e., work begins at $W(0)=0$ and scales quadratically with $\lambda$, as predicted. Expansion of the efficiency gives
\begin{equation}
    \eta(\lambda) = \frac{1}{2} - \frac{\lambda^2}{8\omega^2}+\mathcal{O}(\lambda^4),
\end{equation}
which goes to $1/2$ as $\lambda\rightarrow0$. The efficiency drops to $0$ in the limit $\lambda\rightarrow\infty$, confirming our predictions. This equation does not contradict Eq.~\eqref{eq:eta_small_lambda} since $\Delta'''(0)=0$.

Let us now study the asymptotic limit of work with increased coupling. We have
\begin{equation} \label{eq:W_2qubits_asymptote}
    \lim_{\lambda\rightarrow\infty} W(\lambda) = \omega,
\end{equation}
implying that the work saturates with increased coupling, as predicted.

\begin{figure}
    \centering
    \includegraphics[width=0.5\linewidth]{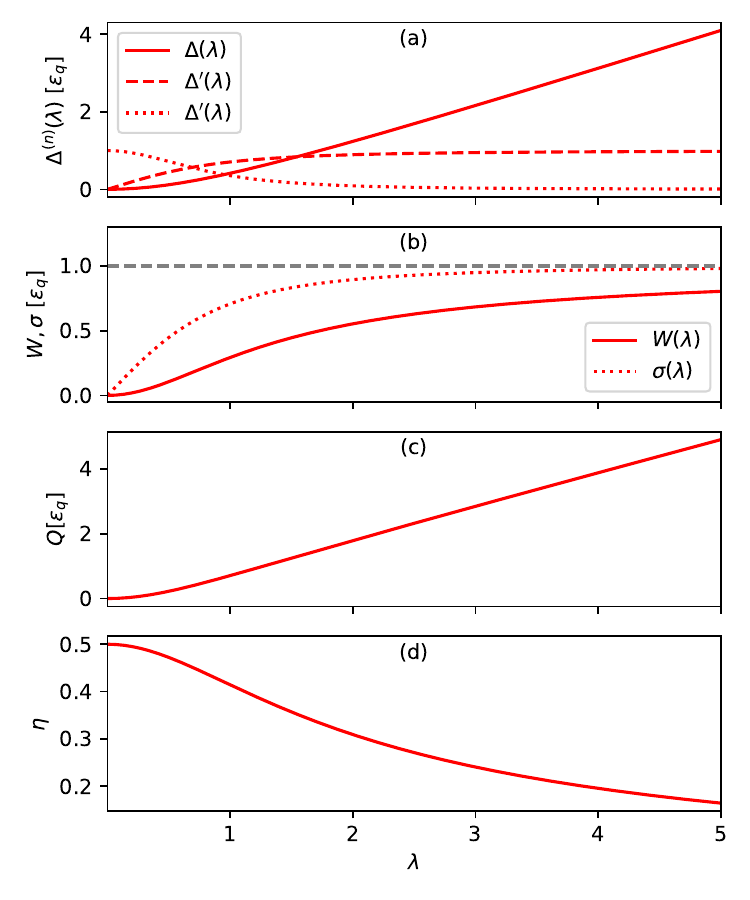}
    \caption{Results describing $N=2$ coupled qubits. (a) QVBF and its derivatives. (b) Work (solid lines) and fluctuations (dotted lines) evaluated from $\sqrt{\sigma^2}$. The grey dashed line corresponds to the asymptotic limit given in Eq.~\eqref{eq:W_2qubits_asymptote} (c) Quantum heat. (d) Efficiency. Panels (a), (b) and (c) are plotted in the units of qubit energy gap $\varepsilon_q$. }
    \label{fig:N_2_qubits}
\end{figure}

In Fig.~\ref{fig:N_2_qubits}, we show the results produced by expressions \eqref{eq:w_2qubits}-\eqref{eq:sigma_2qubits}.

\subsection{Transverse Field Ising Model}

We consider a one-dimensional chain of qubits with nearest-neighbor coupling, described by the Hamiltonian
\begin{equation}
    H(\lambda) = H_\text{loc} + \lambda H_\text{int},
\end{equation}
where the local and interaction parts are given by
\begin{equation}
    H_\text{loc} = \omega\sum_{j=1}^N\sigma_j^\dagger\sigma_j, \quad H_\text{int} = \frac{1}{2}\sum_{j=1}^N\sigma_j^x\sigma_{j+1}^x.
\end{equation}
This model was solved analytically by Juissiau \textit{et al.}; here we verify that their results are recovered within the QVBF framework developed in this work. Juissiau \textit{et al.} derived the following expression for QVBF
\begin{equation}
    \Delta(\lambda) = \frac{1}{2}\left(\sum_p\Omega_p-N\omega\right),
\end{equation}
where the sum runs over fermionic momentum modes
\begin{equation}
    \Omega_p(\lambda)=\sqrt{\omega^2+\lambda^2+2\omega\lambda\cos p}.
\end{equation}
Differentiating this expression yields
\begin{equation} \label{eq:dDelta_TFIM}
    \Delta'(\lambda)=\frac{1}{2}\sum_p\Omega'_p(\lambda) = \frac{1}{2}\sum_p \frac{\lambda + \omega\cos p}{\Omega_p(\lambda)},
\end{equation}
resulting in quantum heat
\begin{equation}
    Q(\lambda) = \lambda\Delta'(\lambda) = \lambda\sum_p \frac{\lambda + \omega\cos p}{2\Omega_p(\lambda)}.
\end{equation}
Using the definition $W(\lambda) = \lambda\Delta'(\lambda)-\Delta(\lambda)$, we obtain
\begin{equation}
    W(\lambda)=\frac{N\omega}{2}-\frac{\omega}{2}\sum_p\frac{\omega+\lambda\cos p}{\Omega_p},
\end{equation}
in agreement with the expression reported by Juissiau \textit{et al.}

\begin{figure}
    \centering
    \includegraphics[width=0.5\linewidth]{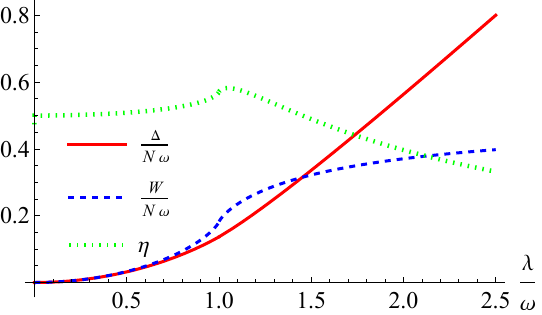}
    \caption{Results for $\Delta, W, \eta$ as a function of $\lambda$, all computed with knowledge of $\Delta(\lambda)$ only.}
    \label{fig:spinchain}
\end{figure}

In the thermodynamic limit, $N \rightarrow \infty$, we can evaluate this result for $\Delta(\lambda)$ by approximating the sum over $p = 2m \pi/N$ where $m = -N/2, \ldots, N/2$ with an integral,
\be
\Delta(\lambda) = \frac{N}{4\pi} \int_{-\pi}^{\pi} dx \sqrt{\omega^2 + \lambda^2 + 2 \omega \lambda \cos x } - \frac{\omega N}{2}.
\ee
We note the even versus odd number of sites does not matter in the thermodynamic limit.
This integral can be evaluated to find
\be
\Delta(\lambda) = \frac{N}{\pi} |\lambda - \omega| {\cal E}\left(- \frac{4 \lambda \omega}{(\lambda-\omega)^2}\right) -  \frac{\omega N}{2},
\ee
where ${\cal E}[\cdot]$ is the complete Elliptic integral of the second kind.  The nonanalytic feature in QVBF is a signature of a ferromagnetic to anti-ferromagnetic phase transition, discussed by Juissau \textit{et al.}
From this function the work, quantum heat and efficiency can be immediately found as in the main text.  They are plotted in Fig.~\ref{fig:spinchain}.

To compute the work fluctuations, both $\bar{e}(\lambda)$ and $\Delta''(\lambda)$ are required. While the latter can be obtained directly by differentiating $\Delta'(\lambda)$, the former depends on the couplings between the ground state and excited states induced by $H_\text{int}$. To evaluate these couplings, we express the interaction Hamiltonian in momentum space and perform a Bogoliubov transformation, following the approach of Jussiau {\it et al}~\cite{jussiau2023many}.

We define the operators
\begin{equation}\label{eq:c_operators_1}
    \hat{c}_j = \exp\left( i\pi \sum_{k=1}^{j-1}\sigma_k^+\sigma_k^-\right)\sigma_j^-,
\end{equation}
followed by their transformation to momentum space
\begin{equation} \label{eq:c_operators_2}
    \hat{c}_p = \frac{1}{\sqrt{N}} \sum_{j=1}^N\hat{c}_j\exp(-ijp).
\end{equation}
The interaction Hamiltonian in terms of $\hat{c}_p$ operators is
\begin{equation}\label{eq:H_int_temp}
    H_\text{int} = \sum_p (\cos p \ A_p + i\sin p \ B_p)
\end{equation}
where
\begin{equation}
\begin{aligned}
    A_p & = \hat{c}_p^\dagger \hat{c}_p - \hat{c}_{-p}\hat{c}_{-p}^{\dagger},\\
    B_p & = \hat{c}_p^\dagger \hat{c}_{-p}^\dagger - \hat{c}_{-p}\hat{c}_p.
\end{aligned}
\end{equation}
The total Hamiltonian $H(\lambda)$ can be diagonalized by the following transformation
\begin{equation}
    \zeta_p = u_p \hat{c}_p + i v_p\hat{c}_{-p}^\dagger,
\end{equation}
where 
\begin{equation}
\begin{aligned}
    u_p & = \frac{\lambda|\sin p|}{\sqrt{2\Omega_p(\Omega_p-\omega-\lambda\cos p)}}, \\
    v_p & = \frac{\text{sgn}\ p}{\sqrt{2}}\sqrt{1 - \frac{\omega+\lambda \cos p}{\Omega_p}}.
\end{aligned}
\end{equation}
satisfying $u_p^2 + v_p^2=1$, $\{\zeta_p^\dagger, \zeta_q\} = \delta_{p,q}$, $\{\zeta_p, \zeta_q\} = 0$. For clarity of the notation, we omit the dependence on $\lambda$ in $\Omega_p(\lambda)$. It leads to the diagonal form of the total Hamiltonian
\begin{equation}
    H(\lambda) = \sum_p\Omega_p\left( \zeta_p^\dagger\zeta_p - \frac{1}{2} \right) + \frac{N\omega}{2},
\end{equation}
which depends on $\lambda$ through $\Omega_p$ and $\zeta_p$ operators. 

To compute $\bar{e}$, we require the interaction Hamiltonian $H_\text{int}$ expressed in terms of the Bogoliubov quasiparticle operators $\zeta_p$. We therefore introduce the number-like operator
\begin{equation}
    N_p = \zeta_p^\dagger \zeta_p + \zeta_{-p}^\dagger \zeta_{-p} - 1,
\end{equation}
and pairwise creation-destruction operator
\begin{equation}
    P_p = \zeta_p^\dagger \zeta_{-p}^\dagger - \zeta_p \zeta_{-p}.
\end{equation}
In terms of these operators, the quantities $A_p$ and $B_p$ can be written as
\begin{equation}
\begin{aligned}
    A_p & = \left( u_p^2 - v_p^2 \right)N_p - 2iu_pv_pP_p,    \\
    B_p & = \left( u_p^2 - v_p^2 \right)P_p - 2iu_pv_p N_p.
\end{aligned}
\end{equation}
Substituting these into \eqref{eq:H_int_temp} yields
\begin{equation}
    H_\text{int} = \sum_p \big\{
    \left[ \left(u_p^2-v_p^2\right) \cos p  + 2u_pv_p \sin p \right] N_p 
    + i \left[ \left(u_p^2-v_p^2\right) \sin p - 2u_pv_p \cos p \right] P_p \big\}.
\end{equation}
Using the identities
\begin{equation}
    u_p^2-v_p^2 = \frac{\omega+\lambda \cos p}{\Omega_p}, \quad 2u_pv_p = \frac{\lambda \sin p}{\Omega_p},
\end{equation}
the interaction Hamiltonian can be simplified to
\begin{equation}
    H_\text{int} = \sum_p\left[ \frac{\lambda + \omega \cos p}{\Omega_p}N_p + i \frac{\omega \sin p}{\Omega_p}P_p \right].
\end{equation}
Expressing this result explicitly in terms of creation and annihilation operators in momentum space gives
\begin{equation}
\begin{aligned}
    H_\text{int} = \sum_p & \bigg[ \frac{\lambda + \omega \cos p}{\Omega_p} \left( \zeta_p^\dagger \zeta_p + \zeta_{-p}^\dagger \zeta_{-p} - 1 \right) \\
    & + i  \frac{\omega \sin p}{\Omega_p} \left( \zeta_p^\dagger \zeta_{-p}^\dagger - \zeta_p \zeta_{-p} \right) \bigg]
\end{aligned}
\end{equation}
This representation makes explicit which excited states are coupled to the ground state by $H_\text{int}$, thereby allowing us to evaluate the coefficients defined in \eqref{eq:cn_en}. We note a minor notational subtlety: following the notation of Juissiau \textit{et al.}, the operators introduced in Eqs.~\eqref{eq:c_operators_1}-\eqref{eq:c_operators_2} are denoted by $\hat{c}_{j(p)}$, whereas the coupling coefficients appearing in Eq.~\eqref{eq:cn_en} are denoted by $c_n$, are written as $c_n$ without a hat. We emphasize this distinction to avoid confusion.

The ground state $|0(\lambda)\rangle$ corresponds to all momentum-space modes occupying their respective ground states. The only term in $H_\text{int}$ that couples the ground state to excited states is proportional to $\zeta_p^\dagger \zeta_{-p}^\dagger$, which creates a pair of excitations with opposite momenta. We denote the resulting state by $|p,-p;\lambda\rangle$, defined through
\begin{equation}
    \zeta_p^\dagger \zeta_{-p}^\dagger |0(\lambda)\rangle = |p,-p;\lambda\rangle. 
\end{equation}
The corresponding coupling coefficients entering \eqref{eq:cn_en} are
\begin{equation}
    c_p^2 = |\langle -p,p;\lambda|H_\text{int}|0(\lambda)\rangle|^2 = \frac{\omega^2 \sin^2 p}{\Omega_p^2}.
\end{equation}
Using these coefficients, the effective excitation energy defined in Eq.~\eqref{eq:1_bar_e} evaluates to
\begin{equation}
    \bar{e}(\lambda) = 2\cfrac{\sum_p \cfrac{\sin^2p}{\Omega_p^2}}{\sum_p \cfrac{\sin^2 p}{\Omega_p^3}}.
\end{equation}
The final ingredient required for computing the work fluctuations is the second derivative of QVBF. Evaluating $\Delta''(\lambda)$ directly from \eqref{eq:dDelta_TFIM} without invoking the thermodynamic limit, yields
\begin{equation}
    \Delta''(\lambda) = \frac{\omega^2}{2}\sum_p \frac{\sin^2 p}{\Omega_p^3}.
\end{equation}
Combining these results according to \eqref{eq:fluctuations}, we obtain
\begin{equation}
    \sigma^2(\lambda) = \frac{\lambda^2\omega^2}{2}\sum_p \frac{\sin^2 p}{\Omega_p^2},
\end{equation}
in agreement with the result derived by Juissiau \textit{et al.} for the transverse-field Ising model.  In the thermodynamic limit, the sum can be replaced by an integral as before to find
\be
\sigma^2 = \frac{N}{2} \begin{cases}
    \lambda^2, & \lambda \le \omega \\
    \omega^2, & \lambda > \omega.
\end{cases}
\ee

\subsection{Qubits with random couplings}

To verify our results for a broad class of Hamiltonians beyond analytically solvable models, we implement a numerical framework that evaluates all relevant thermodynamic quantities both using the expressions proposed by Juissiau \textit{et al.} and within the QVBF formalism developed in this manuscript. The engines are evaluated for Hamiltonians of the form
\begin{equation}
    H(\lambda) = H_\text{loc} + \lambda H_\text{int},
\end{equation}
with 
\begin{equation}
    H_\text{loc} = \frac{1}{2}\sum_{j=1}^N \sigma_z^{(j)}, H_\text{int} = \frac{1}{2}\sum_{j=1}^N\sum_{k=j+1}^N g_{jk} \sigma_x^{(j)} \sigma_x^{(k)},
\end{equation}
where the coupling coefficients $g_{jk}$ may either be specified explicitly or sampled randomly. 

In the numerical examples presented in this manuscript, the couplings ($g_{jk}$) are drawn independently from a uniform distribution on the interval $[-1,1]$ and subsequently normalized such that the $l2$ norm norm of the vector containing all couplings satisfies $\sqrt{\sum_{j<k}g_{jk}^2}=1$. This normalization ensures that the overall interaction strength remains fixed as the number of qubits is varied. 

The numerical framework based on QVBF was implemented in \textsc{PyTorch} v.~2.7.0. All derivatives $\Delta^{(n)}(\lambda)$ are obtained via automatic differentiation.

The described framework was used to produce the $10$-qubit example presented in the main body of this manuscript. The couplings between qubits used to produce these results are given in Tab.~\ref{tab:couplings}.

\begin{table}[h] 
\centering
\begin{tabular}{cc}
\textbf{Coupling} & \textbf{Value} \\
\hline
$g_{1,2}$ & $-0.06217543$ \\
\hline
$g_{1,3}$ & $0.22336509$ \\
\hline
$g_{1,4}$ & $0.11497161$ \\
\hline
$g_{1,5}$ & $0.04889319$ \\
\hline
$g_{1,6}$ & $-0.17047036$ \\
\hline
$g_{1,7}$ & $-0.17048231$ \\
\hline
$g_{1,8}$ & $-0.21900502$ \\
\hline
$g_{1,9}$ & $0.18146965$ \\
\hline
$g_{1,10}$ & $0.05011060$ \\
\hline

$g_{2,3}$ & $0.10311665$ \\
\hline
$g_{2,4}$ & $-0.23758884$ \\
\hline
$g_{2,5}$ & $0.23287803$ \\
\hline
$g_{2,6}$ & $0.16475200$ \\
\hline
$g_{2,7}$ & $-0.14255905$ \\
\hline
$g_{2,8}$ & $-0.15768125$ \\
\hline
$g_{2,9}$ & $-0.15689846$ \\
\hline
$g_{2,10}$ & $-0.09701367$ \\
\hline
$g_{3,4}$ & $0.01226880$ \\
\hline
$g_{3,5}$ & $-0.03372670$ \\
\hline
$g_{3,6}$ & $-0.10346271$ \\
\hline
$g_{3,7}$ & $0.05543208$ \\ \hline
$g_{3,8}$ & $-0.17865971$ \\ \hline
$g_{3,9}$ & $-0.10300900$ \\ \hline
$g_{3,10}$ & $-0.06622843$ \\ \hline

$g_{4,5}$ & $-0.02177085$ \\ \hline
$g_{4,6}$ & $0.14132756$ \\ \hline
$g_{4,7}$ & $-0.14883573$ \\ \hline
$g_{4,8}$ & $0.00705431$ \\ \hline
$g_{4,9}$ & $0.04579883$ \\ \hline
$g_{4,10}$ & $-0.22477020$ \\ \hline

$g_{5,6}$ & $0.05329710$ \\ \hline
$g_{5,7}$ & $-0.16328173$ \\ \hline
$g_{5,8}$ & $-0.21555183$ \\ \hline
$g_{5,9}$ & $0.22245879$ \\ \hline
$g_{5,10}$ & $0.23075802$ \\ \hline

$g_{6,7}$ & $0.15283562$ \\ \hline
$g_{6,8}$ & $-0.09682955$ \\ \hline
$g_{6,9}$ & $-0.19938574$ \\ \hline
$g_{6,10}$ & $0.09130224$ \\ \hline

$g_{7,8}$ & $-0.02965924$ \\ \hline
$g_{7,9}$ & $-0.18731037$ \\ \hline
$g_{7,10}$ & $-0.00239023$ \\ \hline

$g_{8,9}$ & $-0.23074784$ \\ \hline
$g_{8,10}$ & $0.20285109$ \\ \hline

$g_{9,10}$ & $-0.11954387$ \\ \hline
\end{tabular}
\caption{Coupling coefficients $g_{ij}$ used to generate the numerical results describing $10$-qubit model discussed in the main body of the paper.}\label{tab:couplings}
\end{table}

\subsection{Two coupled oscillators - model 1}

In this section, we consider a system of two coupled oscillators with the Hamiltonian
\begin{equation}
    H = \tfrac{1}{2}\left(p_1^2+p_2^2\right) + \tfrac{k_0}{2}\left( x_1^2+x_2^2 \right) + \frac{g}{2}\left( x_1-x_2\right)^2,
\end{equation}
which can be divided into following parts
\begin{equation}
    H_\text{loc} = \tfrac{1}{2}\left(p_1^2+p_2^2\right) + \tfrac{1}{2}\left(  k_0+g\right)\left( x_1^2+x_2^2\right),
\end{equation}
\begin{equation} \label{eq:H_int_2_oscillators}
    H_\text{int} = - g x_1 x_2.
\end{equation}
This model was considered by Juissau \textit{et al.} We aim to verify if the theory developed in this manuscript recovers their results. 

We begin by noticing that the parameter $g$ appears in both $H_\text{loc}$ and $H_\text{int}$. Thus, we cannot identify $g$ with $\lambda$. To make our theories agree, we introduce $\lambda$ as an additional parameter and write
\begin{equation}
    H(\lambda) = H_\text{loc} + \lambda H_\text{int},
\end{equation}
and taking $\lambda=1$ should recover the results given by Juissau et. al. Just as in their paper, we introduce
\begin{equation} \label{eq:x_pm}
    x_\pm = \tfrac{1}{\sqrt{2}}(x_1\pm x_2), \quad p_\pm = \tfrac{1}{\sqrt{2}}(p_1\pm p_2),
\end{equation}
and write the Hamiltonian in terms of these variables. This results in Hamiltonian describing two decoupled oscillators
\begin{equation}
    H(\lambda) = \sum_{j=+,-} \left[\tfrac{1}{2}p_j^2 + \tfrac{1}{2}\omega_j(\lambda)^2x_j^2 \right],
\end{equation}
where
\begin{equation}
    \omega_\pm = \sqrt{k_0+(1\mp\lambda)g}.
\end{equation}
The ground state of this Hamiltonian is composed of the ground states of both oscillators
\begin{equation}
    E_0(\lambda) = \tfrac{1}{2}\left[ \omega_+(\lambda) + \omega_-(\lambda) \right],
\end{equation}
which implies the QVBF
\begin{equation} \label{eq:delta_2_oscillators}
\begin{aligned}
    \Delta(\lambda) & = \sqrt{k_0 + g} \\ & - \tfrac{1}{2}\sqrt{k_0+g-\lambda g} - \tfrac{1}{2}\sqrt{k_0+g+\lambda g}    \\
    & = \omega - \frac{\omega_+(\lambda)+\omega_-(\lambda)}{2},
\end{aligned}
\end{equation}
where we used
\begin{equation}
    \omega\equiv\sqrt{k_0+g}.
\end{equation}
The derivative of QVBF is
\begin{equation}
\begin{aligned}
    \Delta'(\lambda) & = -\frac{1}{2}\left[ \omega_+'(\lambda) - \omega_-(\lambda)\right] \\
    & = - \frac{g}{4}\left[\frac{1}{\omega_-(\lambda)} - \frac{1}{\omega_+(\lambda)} \right].
\end{aligned}
\end{equation}
This allows us to compute the quantum heat, which after some algebra yields
\begin{equation}
\begin{aligned}
    Q(\lambda) & = \lambda \Delta'(\lambda) \\
    & = \frac{\omega_+(\lambda)+\omega_-(\lambda)}{4} \left( \frac{\omega^2}{\omega_+(\lambda) \ \omega_-(\lambda)} - 1\right).
\end{aligned}
\end{equation}
It agrees with the expression provided by Juissau \textit{et al.} if we make the replacement $\omega_\pm\rightarrow\omega_\pm(\lambda)$. Taking $\lambda=1$ recovers their result. Combining it with \eqref{eq:delta_2_oscillators} results in
\begin{equation}
    W(\lambda) = \frac{\omega_+(\lambda)+\omega_-(\lambda)}{4} \left( \frac{\omega^2}{\omega_+(\lambda) \ \omega_-(\lambda)} + 1\right) - \omega,
\end{equation}
which agrees with the result of Juissau \textit{et al.} if we take $\lambda=1$. Let us now compute the fluctuations. We will need the second derivative
\begin{equation}
\begin{aligned}
    \Delta''(\lambda) & = -\tfrac{1}{2}\left[ \omega_+''(\lambda) + \omega_-''(\lambda)  \right] \\
    &  = \frac{g^2}{8} \left[ \omega_+(\lambda)^{-3} + \omega_-(\lambda)^{-3}\right]
\end{aligned}
\end{equation}
The other ingredient needed is the effective energy $\bar{e}(\lambda)$. To compute it, we will work with the interaction Hamiltonian in terms of creation an annihilation operators describing the decoupled oscillators
\begin{equation} \label{eq:a_pm}
    x_\pm  = \frac{a_\pm^\dagger+a_\pm}{\sqrt{2\omega_\pm(\lambda)}}. 
\end{equation}
$H_\text{int}$ contains the squares of these operators, implying that it couples only the states differing by two excitations. It couples the ground state of both oscillators $|0_+,0_-\rangle$ only to two other states, and the only non-zero $c_n$ coefficiencts defined in \eqref{eq:cn_en} are
\begin{equation} \label{eq:c_plus}
    c_+=|\langle2_+,0_-|H_\text{int}|0_+,0_-\rangle|^2 = \frac{g^2}{8\omega_+^2(\lambda)},
\end{equation}
and
\begin{equation} \label{eq:c_minus}
    c_- = |\langle0_+,2_-|H_\text{int}|0_+,0_-\rangle|^2 = \frac{g^2}{8\omega_-^2(\lambda)},
\end{equation}
with corresponding energy differences
\begin{equation} \label{eq:e_pm}
    e_+=2\omega_+(\lambda), \quad e_-=2\omega_-(\lambda).
\end{equation}
The effective excitation energy is given by
\begin{equation}
    \bar{e}(\lambda) = 2 \frac{\omega_+(\lambda)^{-2}+\omega_-(\lambda)^{-2}}{\omega_+(\lambda)^{-3}+\omega_-(\lambda)^{-3}}.
\end{equation}
Calculating the fluctuations from \eqref{eq:fluctuations} yields
\begin{equation}
    \sigma^2(\lambda) = \frac{\lambda^2g^2\omega^2}{4\omega_+(\lambda)^2\omega_-(\lambda)^2}.
\end{equation}
Let us not take the square root and evaluate it at $\lambda=1$
\begin{equation}
    \sigma(1) = \frac{g\omega}{2 \omega_+(1)^2\omega_-(1)^2},
\end{equation}
which agrees with the expression provided by Juissau \textit{et al.} This shows that our predictions are consistent with each other.

\subsection{Two coupled oscillators - model 2}

In this section, we consider an alternative version of the engine consisting of two coupled harmonic oscillators, where the interaction strength is directly controlled by the parameter $\lambda$. The Hamiltonian is given by
\begin{equation}
    H(\lambda) = \frac{p_1^2}{2} + \frac{k x_1^2}{2} + \frac{p_2^2}{2} + \frac{k x_2^2}{2} + \lambda x_1 x_2,
\end{equation}
where the interaction part is 
\begin{equation} \label{eq:2_oscillators_x1_x2}
    H_\text{int} = x_1x_2.
\end{equation}

In contrast to the model discussed previously, varying $\lambda$ here directly tunes the interaction strength, making it a natural physical control parameter of the engine.

To diagonalize the Hamiltonian, we introduce the normal-mode coordinates $x_\pm$ defined in \eqref{eq:x_pm}, which yields
\begin{eqnarray}
    H(\lambda) = \sum_{j=+,-} \left[ \frac{p_j^2}{2} + \frac{1}{2}\omega_j^2x_j^2 \right],
\end{eqnarray}
with frequencies
\begin{equation} \label{eq:omega_pm_model2}
    \omega_\pm(\lambda) = \sqrt{k\pm\lambda}.
\end{equation}
We observe that one of these frequencies becomes imaginary when $|\lambda|>k$, indicating the instability of the system. Consequently, the analysis that follows is restricted to the stable regime $|\lambda|<k$. 

The ground-state energy of the coupled oscillators is
\begin{eqnarray}
    E_0(\lambda) = \frac{1}{2}[\omega_+(\lambda) + \omega_-(\lambda)],
\end{eqnarray}
which leads to QVBF
\begin{eqnarray}
    \Delta(\lambda) = \sqrt{k} - \frac{1}{2}\left(\sqrt{k+\lambda} + \sqrt{k-\lambda}\right),
\end{eqnarray}
with derivatives
\begin{eqnarray}
\begin{aligned}
    \Delta'(\lambda) & = \frac{1}{4}\left( \frac{1}{\sqrt{k-\lambda}} - \frac{1}{\sqrt{k+\lambda}} \right), \\
    \Delta''(\lambda) & = \frac{1}{8}\left( \frac{1}{(k+\lambda)^{3/2}} + \frac{1}{(k-\lambda)^{3/2}} \right).
\end{aligned}
\end{eqnarray}
As expected, $\Delta''(\lambda)>0$ for $|\lambda|\leq k$ within the stable regime. 

We obtain the quantum heat from the first derivative of the QVBF 
\begin{equation}
    Q(\lambda) = \frac{\lambda}{4}\left( \frac{1}{\sqrt{k-\lambda}} - \frac{1}{\sqrt{k+\lambda}} \right),
\end{equation}
and work
\begin{equation}
\begin{aligned}
    W(\lambda) = \frac{\lambda}{4}\left( \frac{1}{\sqrt{k-\lambda}} - \frac{1}{\sqrt{k+\lambda}} \right) - \sqrt{k} + \frac{1}{2}(\sqrt{k+\lambda} + \sqrt{k-\lambda}).
\end{aligned}
\end{equation}
We now turn to the computation of work fluctuations. The interaction hamiltonian can be expressed as
\begin{equation}
    H_\text{int} = \frac{x_+^2-x_-^2}{2},
\end{equation}
which is quadratic in position operators. Together with Eq.~\eqref{eq:a_pm}, this form implies that $H_\text{int}$ couples the ground state $|0(\lambda)\rangle=|0_+,0_-\rangle$ only to two-excitation states $|2_+,0_-\rangle$ and $|0_+,2_-\rangle$, in close analogy with the previous oscillator model. 

The coupling coefficients defined in Eq.~\eqref{eq:cn_en} are
\begin{equation}
\begin{aligned}
    c_+&=|\langle2_+,0_-|H_\text{int}|0_+,0_-\rangle|^2 = \frac{1}{8\omega_+^2(\lambda)}, \\
    c_-&=|\langle0_+,2_-|H_\text{int}|0_+,0_-\rangle|^2 = \frac{1}{8\omega_-^2(\lambda)}, 
\end{aligned}
\end{equation}
with associated excitation energies $e_\pm=2\omega_\pm(\lambda)$. This leads to the effective coupling energy
\begin{eqnarray}
    \bar{e}(\lambda) = 2 \frac{\omega_+(\lambda)^{-2}+\omega_-(\lambda)^{-2}}{\omega_+(\lambda)^{-3}+\omega_-(\lambda)^{-3}}.
\end{eqnarray}
Substituting these results into the general expression for the work fluctuations yields
\begin{equation}
    \sigma^2(\lambda) = \frac{\lambda^2 k}{4(k^2 - \lambda^2)}.
\end{equation}
The fluctuations are positive throughout the stable regime $|\lambda|<k$, diverge at $|\lambda|=k$, This behavior provides a complementary perspective on the stability of the engine, reinforcing the conclusion that physically meaningful operation is restricted to $|\lambda|<k$.

\subsection{Oscillator Chain}
\begin{figure}
    \centering
    \includegraphics[width=0.5\linewidth]{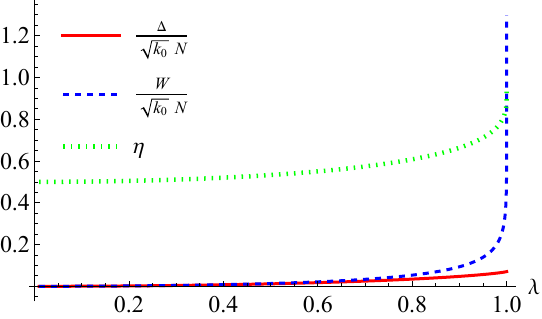}
    \caption{For the oscillator chain, we plot the scaled QVBF, work and efficiency of the engine versus the interaction strength $\lambda$.  The logarithmic divergence at $\lambda =1$ causes the efficiency to limit to unity and the work to diverge.  }
    \label{fig:oscillatorchain}
\end{figure}

It is also of interest to examine the oscillator chain.  The Hamiltonian is of the form
\be
H = \sum_{j=1}^N p_j^2/2 + \frac{1}{2} \sum_{jk} x_j K_{jk} x_k.
\ee
We consider the oscillator coupling matrix to be of tridiagonal form
\be
K = k_0 (2 I - \lambda T)
\ee
where $I$ is the identity matrix, and $T$ has 1s on the first off-diagonals and 0s on the main diagonal.  The quantum system can be solved exactly with a Gaussian form ground state wavefunction.  Jussiau {\it et al.}~\cite{jussiau2023many} found that the QVBF is given by
\be
\Delta = \frac{1}{2} \sum_{j=1}^N [\sqrt{K_{jj}} - (\sqrt{K})_{jj}].
\ee
The solution of the oscillator chain can be adapted to the parameter $\lambda$ to find the eigenvalues of the matrix $K$ are given by $k_j = 2 k_0 (1-\lambda \cos(j \pi/(N+1))$.  In this model of equal couplings, the QVBF is given by
\be
\Delta(\lambda) = \frac{1}{2} \sum_j (\sqrt{2 k_0} - \sqrt{2 k_0 \left(1-\lambda \cos(\frac{\pi j}{N+1})\right)}.
\ee
In the large $N$ thermodynamic limit, we can approximate the sum with an integral, to find
\be
\Delta(\lambda) = \frac{N}{\pi} \sqrt{k_0/2} \int_0^\pi dx (1 - \sqrt{1-\lambda \cos x}).
\ee
The integral is given in analytic form by
\be
\Delta(\lambda) = N \sqrt{k_0/2} \left(1 - \frac{2\sqrt{1+\lambda}}{\pi} {\cal E}\left(\frac{2\lambda}{1+\lambda}\right)\right),
\ee
where once again, ${\cal E}[\cdot]$ is the complete Elliptic integral of the second kind.  The scaled QVBF $\Delta(\lambda)$, work $W(\lambda) = \lambda \Delta' - \Delta$ and efficiency $\eta(\lambda) = 1 - \Delta/(\lambda \Delta')$ are plotted in Fig.~\ref{fig:oscillatorchain}.   There we see that at $\lambda = 1$, a divergence of the work occurs, resulting in the efficiency to limit to 1.  This feature originates in the behavior of the Elliptic integral of the second kind - while the function itself remains finite and continuous (as can be seen in the behavior of $\Delta$), its derivative has a logarithmic divergence, signaling the break down of the theory at this point.  A more careful analysis at the point $\lambda=1$  by Jussiau {\it et al.}~\cite{jussiau2023many} shows that the work scales as $N \ln N$, while the quantum heat still scales linearly in $N$, reflecting this singularity in the theory.  This behavior results in near perfect efficiency of the engine, in principle.

The fluctuation's dependence on $\lambda$ can be calculated in a similar manner.  The fluctuations are given in Jussiau {\it et al.}~\cite{jussiau2023many} as
\begin{eqnarray}
\sigma^2 &=& \frac{k_0^2}{2} {\rm Tr} K^{-1} - N k_0/4 \\
&=& \frac{k_0}{2} \sum_{j=1}^N \frac{1}{(1-\lambda \cos(j \pi /(N+1))} - 1\\
&\approx& \frac{k_0 N}{4} \left(\int_0^\pi \frac{dx}{\pi} \frac{1}{1-\lambda \cos x} -1\right).
\end{eqnarray}
Evaluating the integral, we find the result
\be
\sigma^2 = \frac{k_0 N}{4}\left(\frac{1}{\sqrt{1-\lambda^2} } -1\right).
\ee
We see that the fluctuation diverge at the point $\lambda = \pm 1$, coinciding at the point when the efficiency limits to 1.  A more careful evaluation of the sum indicated the scaling law $\sigma^2 \sim N^2$ at this point, showing much larger fluctuations at this point.

\subsection{Single qubit with external field}

We consider a single qubit engine with an external field coupling to the $\sigma_x$ operator
\begin{equation}
    H(\lambda) = \frac{1}{2}\sigma_z + \frac{\lambda}{2}\sigma_x.
\end{equation}
The eigenvalues are
\begin{equation}
    E_0(\lambda) = - \tfrac{1}{2}\sqrt{1+\lambda^2},\quad E_1(\lambda) = \tfrac{1}{2}\sqrt{1+\lambda^2}.
\end{equation}
QVBF is
\begin{equation}
    \Delta(\lambda) = -\tfrac{1}{2} + \tfrac{1}{2}\sqrt{1+\lambda^2},
\end{equation}
with derivatives
\begin{equation}
    \Delta'(\lambda) = \frac{\lambda}{2\sqrt{1+\lambda^2}}, \quad \Delta''(\lambda) = \frac{1}{2\left(1+\lambda^2\right)^{3/2}}.
\end{equation}
Quantum heat becomes
\begin{equation}
    Q(\lambda) =\lambda\Delta'(\lambda)= \frac{\lambda^2}{2\sqrt{1+\lambda^2}}
\end{equation}
and work evaluates to 
\begin{equation}
    W(\lambda) = Q(\lambda)-\Delta(\lambda)=\frac{1}{2}-\frac{1}{2\sqrt{1+\lambda^2}},
\end{equation}
resulting in efficiency
\begin{equation}
    \eta(\lambda) = \frac{W(\lambda)}{Q(\lambda)} = \frac{\sqrt{1+\lambda^2}-1}{\lambda^2}. 
\end{equation}
Since the engine has only one excited state, we have 
\begin{equation}
    \bar{e}(\lambda)=E_1(\lambda)-E_0(\lambda)=\sqrt{1+\lambda^2},
\end{equation}
allowing us to compute the fluctuations
\begin{equation}
    \sigma^2(\lambda) = \tfrac{1}{2}\lambda^2\Delta''(\lambda) \bar{e}(\lambda) = \frac{\lambda^2}{4\left(1+\lambda^2\right)}.
\end{equation}

\subsection{Harmonic oscillator with external field}

We consider a single-body engine consisting of a harmonic oscillator in external field. The Hamiltonian is given by
\begin{equation}
\begin{aligned}
    H(\lambda) & = \frac{p^2}{2} + \frac{x^2}{2} + \lambda x \\
    & =  \frac{p^2}{2} + \frac{\left(x +\lambda \right)^2}{2} - \frac{\lambda^2}{2},
\end{aligned}
\end{equation}
which we can split into the local and interaction part
\begin{equation}
    H_\text{loc} = \frac{p^2}{2} + \frac{x^2}{2}, \quad H_\text{int} = x.
\end{equation}
The external field displaces the equilibrium position of the oscillator. The ground state energy is
\begin{equation}
    E(\lambda) = \frac{1}{2} - \frac{\lambda^2}{2},
\end{equation}
resulting in QVBF
\begin{equation}
    \Delta(\lambda) = \frac{\lambda^2}{2},
\end{equation}
with derivatives
\begin{equation}
    \Delta'(\lambda) = \lambda, \quad \Delta''(\lambda)=1.
\end{equation}
This implies quantum heat
\begin{equation} \label{eq:heat_oscillator_field}
    Q(\lambda) = \lambda \Delta'(\lambda) = \lambda^2,
\end{equation}
work
\begin{equation} \label{eq:work_oscillator_field}
    W(\lambda) = Q(\lambda) - \Delta(\lambda) = \frac{\lambda^2}{2},
\end{equation}
and efficiency
\begin{equation}
    \eta(\lambda) = \frac{W(\lambda)}{Q(\lambda)} = \frac{1}{2}.
\end{equation}
Let us check if this theory is consistent with the generalized version of expressions provided by Juissau \textit{et al.} We need the entangled ground state which is the coherent state
\begin{equation}
    |\alpha=\tfrac{\lambda}{\sqrt{2}}\rangle={}{}|0(\lambda)\rangle = D\left( \frac{\lambda}{\sqrt{2}} \right) |0(0)\rangle,
\end{equation}
where a coherent state must satisfy $|\alpha\rangle =D(\alpha)|0\rangle$, and $a|\alpha\rangle=\alpha|\alpha\rangle$.
The expected value of $H_\text{int}$ in that state is
\begin{equation}
    \langle 0(\lambda)|H_\text{int}|0(\lambda)\rangle=\frac{\lambda^2}{2}+\frac{1}{2}.
\end{equation}
Generalization of expression of work provided by Juissau \textit{et al.} to a single-body problem is
\begin{equation}
    W(\lambda) = \langle 0(\lambda)|H_c|0(\lambda)\rangle - E_0(0),
\end{equation}
and evaluating it returns Eq.~\eqref{eq:work_oscillator_field}. Heat can be computed from $Q(\lambda) = W(\lambda)+\Delta(\lambda)$, and evaluating it returns Eq.~\eqref{eq:heat_oscillator_field}. Therefore, efficiency must agree as well. 
The excited states of the oscillator can be obtained through the displacement operator
\begin{equation}
    |n(\lambda)\rangle = D\left( \frac{\lambda}{\sqrt{2}} \right) |n(0)\rangle.
\end{equation}
This helps us with computing the fluctuations. We need the coefficients 
\begin{equation}
\begin{aligned}
    c_n & = |\langle n(\lambda) |x  |0(\lambda)\rangle|^2 \\
    & = |\langle n(0) | D^\dagger\left( \lambda/\sqrt{2} \right)x D\left( \lambda/\sqrt{2} \right) |0(0)\rangle|^2 \\
    & = |\langle n(0) | (x +\lambda) |0(0)\rangle|^2 = |\langle n(0) | x |0(0)\rangle|^2.
\end{aligned}
\end{equation}
The only non-zero coefficient contains coupling to the first excited state with the corresponding energy $\bar{e}=e_1=1$. This yields
\begin{equation}
    \sigma^2(\lambda) = \frac{1}{2}\lambda^2\Delta''(\lambda)\bar{e}(\lambda) = \frac{\lambda^2}{2}.
\end{equation}
This can be computed also from the expression provided by Juissau \textit{et al.} adapted to single-body case 
\begin{equation}
    \sigma^2(\lambda) = \langle H_\text{loc}^2\rangle_\lambda-\langle H_\text{loc} \rangle_\lambda^2,
\end{equation}
which yields the same result.
\putbib[references]
\end{bibunit}
\thispagestyle{empty}
\end{document}